# On the Power of Attribute-based Communication⋆

Yehia Abd Alrahman[1], Rocco De Nicola[1], and
Michele Loreti[2]

[1] IMT Institute for Advanced Studies Lucca, Italy
[2] Università degli Studi di Firenze

**Abstract.** In open systems, i.e. systems operating in an environment that they cannot control and with components that may join or leave, behaviors can arise as side effects of intensive components interaction. Finding ways to understand and design these systems and, most of all, to model the interactions of their components, is a difficult but important endeavor. To tackle these issues, we present *AbC*, a calculus for attribute-based communication. An *AbC* system consists of a set of parallel agents each of which is equipped with a set of attributes. Communication takes place in an implicit multicast fashion, and interactions among agents are dynamically established by taking into account "connections" as determined by predicates over the attributes of agents. First, the syntax and the semantics of the calculus are presented, then expressiveness and effectiveness of AbC are demonstrated both in terms of modeling scenarios featuring collaboration, reconfiguration, and adaptation and of the possibility of encoding channel-based interactions and other interaction patterns. Behavioral equivalences for AbC are introduced for establishing formal relationships between different descriptions of the same system.

## 1 Introduction

In a world of *Internet of Things* (IoT), of *Systems of Systems* (SoS), and of *Collective Adaptive Systems* (CAS), most of the concurrent programming models still rely on communication primitives based on point-to-point, multicast with explicit addressing (i.e. IP multicast [13]), or on broadcast communication. In our view, it is important to consider alternative basic interaction primitives and in this paper we study the impact of a new paradigm that permits selecting groups of partners by considering the (predicates over the) attributes they expose.

The findings we report in this paper have been triggered by our interest in CAS, see e.g. [5], and by the recent attempts to define appropriate abstractions and linguistic primitives to deal with such systems, see e.g. SCEL [7] and the calculus presented in [2].

---

⋆ This research has been supported by the European projects IP 257414 ASCENS and STReP 600708 QUANTICOL, and by the Italian project PRIN 2010LHT4KM CINA.

CAS consists of large numbers of interacting components which exhibit complex behaviors depending on their attributes, objectives and actions. Decision-making in such systems is complex and interaction between components may lead to unexpected behaviors. CAS(s) are open, in that components may enter or leave the collective at anytime and may have different (potentially conflicting) objectives; so they need to dynamically adapt to new requirements and contextual conditions. New engineering techniques to address the challenges of developing, integrating, and deploying such systems are needed [28].

To move towards this goal, in our view, it is important to develop a theoretical foundation for this class of systems that would help in understanding their distinctive features. In this paper, we concentrate our attention on $AbC$, a calculus inspired by SCEL and focus on a minimal set of primitives that permits attribute-based communication. $AbC$ systems are represented as sets of parallel components, each of which is equipped with a set of attributes whose values can be modified by internal actions. Communication actions (both send and receive) are decorated with predicates over attributes that partners have to satisfy to make the interaction possible. Thus, communication takes place in an implicit multicast fashion, and communication partners are selected by relying on predicates over the exposed attributes in their interfaces. Unlike IP multicast [13] where the reference address of the group is explicitly included in the message, $AbC$ components are unaware of the existence of each other and they receive messages only if they satisfy the sender's requirements. The semantics of output actions is non-blocking while input actions are blocking in that they can only take place through synchronization with available sent messages.

Many communication models addressing distributed systems have been introduced so far. Some of the well-known approaches include: channel-based models [17, 12, 19], group-based models [1, 4, 13], and publish/subscribe models [3, 10].

- *Channel-based* models rely on explicit naming of communication partners , in $AbC$ the interacting partners are anonymous to each other. Rather than agreeing on channels or names, they interact by relying on the satisfaction of predicates over the attributes they expose. This makes our calculus more suitable for modeling scalable distributed systems as anonymity is a key factor for scalability.
- In *group-based* models, the formation of the collective is static in the sense that the groups are explicitly specified in advance; in $AbC$, collectives are dynamically formed (the group name is modeled as an attribute) and destroyed at the time of interaction. There is no need for a construct for group formation or destruction.
- The *publish/subscribe* model is just a special case of $AbC$ where publishers can attach attributes to messages and send them with empty predicates (i.e., satisfied by all). Only subscribers can check the compatibility of the attached publishers attributes with their subscriptions.

To further support our approach, we would like to stress that attributes make it easy to encode interesting feature of CAS. For instance, awareness can be easily modeled by locally reading the values of the attributes that represent either the component status (e.g., the battery level of a robot) or the external environment (e.g., the external humidity). Also localities of CAS components can be naturally modeled as attributes. In fact, the general concept of attribute-based communication can be exploited to provide a general unifying framework to encompass different communication models and interaction patterns such as those outlined above and many others.

The $AbC$ calculus presented in this paper is a refined and extended version of the one presented in [2] which from now on we shall call "the old $AbC$". The latter is a very basic calculus with a number of limitations, see the discussion in Section 6. Here, we fully redesign the calculus, enrich it with behavioral equivalences and assess expressiveness and effectiveness of the new proposal. More specifically, the main contributions of this paper are:

1. A new version of $AbC$ calculus featuring replication for modeling open-ended systems, name restriction, multithreading, an awareness operator for acquiring knowledge about both local status and external environment, a richer language for defining predicates over attributes, see Section 2 and Section 3.

2. The study of the expressive power of $AbC$ both in terms of the ability of modeling scenarios featuring collaboration, reconfiguration, and adaptation and of the possibility of encoding channel-based communication and other communication paradigms, see Section 4.

3. The definition of behavioral equivalences for $AbC$ by first introducing a context based reduction barbed congruence relation and then the corresponding extensional labelled bisimilarity, see Section 5.

4. The proof of the correctness of the translation from $b\pi$-calculus [8] into $AbC$ up to the introduced equivalence, see Section 5.3.

In the following sections, the main features of $AbC$ will be presented in a step-by-step fashion using a running example from the swarm robotics domain described below. A complete $AbC$ model of this scenario is given in Section 4.1.

We consider a scenario where a swarm of robots spreads throughout a given disaster area with the goal of locating and rescuing possible victims. All robots playing the same role execute the same code, defining the functional behavior, and a set of adaptation mechanisms, regulating the interactions among robots and their environments. All robots initially play the explorer role to search for victims in the environment. Once a robot finds a victim, it changes its role to "rescuer" and sends victim's information to nearby explorers. The collective (the swarm) starts forming in preparation for the rescuing procedure. As soon as another robot receives victim's information, it changes its role to "helper" and moves to join the rescuers-collective. The rescuing procedure starts only when the collective formation is complete. During exploration, in case of critical battery level, a robot enters a power saving mode until it is recharged.

$$C \quad ::= \Gamma{:}P \quad | \quad C_1 \| C_2 \quad | \quad !C \quad | \quad \nu x C$$

$$P \quad ::= \mathbf{0} \quad | \quad \Pi(\tilde{x}).P \quad | \quad (\tilde{E})@\Pi.P \quad | \quad [\tilde{a} := \tilde{E}]P \quad | \quad \langle \Pi \rangle P \quad | \quad P_1 + P_2 \quad | \quad P_1 | P_2 \quad | \quad K$$

$$\Pi \quad ::= \mathtt{tt} \quad | \quad \mathtt{ff} \quad | \quad E_1 \bowtie E_2 \quad | \quad \Pi_1 \land \Pi_2 \quad | \quad \Pi_1 \lor \Pi_2 \quad | \quad \neg \Pi$$

$$E \quad ::= v \quad | \quad x \quad | \quad a \quad | \quad this.a$$

**Table 1.** The syntax of the *AbC* calculus

## 2   The AbC Calculus

*AbC* aims at modeling highly complex and adaptive systems (e.g., CAS) with the appropriate level of abstraction that permits natural modeling and reasonable verification through compact models. The constructs have been designed to serve this purpose. The brand new constructs model CAS concepts as first class citizens. For instance, they model adaptation through attribute updates, awareness with explicit constructs, anonymous interaction and collective formation by relying on predicates over attributes (instead of explicit channels) to determine partners.

The syntax of the *AbC* calculus is reported in Table 1. The top-level entities of the calculus are *components* ($C$), a component is either a process $P$ associated with an *attribute environment* $\Gamma$, denoted $\Gamma{:}P$, or the parallel composition $C_1 \| C_2$ of two components, or the replicating component $!C$ which can always create a new copy of $C$. The *attribute environment* $\Gamma{:}\mathcal{A} \rightharpoonup \mathcal{V}$

| | |
|---|---|
| $\Gamma \models \mathtt{tt}$ | for all $\Gamma$ |
| $\Gamma \models \mathtt{ff}$ | for no $\Gamma$ |
| $\Gamma \models E_1 \bowtie E_2$ | iff $\Gamma(E_1) \bowtie \Gamma(E_2)$ |
| | where $\Gamma(v) = v$ |
| $\Gamma \models \Pi_1 \land \Pi_2$ | iff $\Gamma \models \Pi_1$ and $\Gamma \models \Pi_2$ |
| $\Gamma \models \Pi_1 \lor \Pi_2$ | iff $\Gamma \models \Pi_1$ or $\Gamma \models \Pi_2$ |
| $\Gamma \models \neg \Pi$ | iff not $\Gamma \models \Pi$ |

**Table 2.** The predicate satisfaction

is a partial map from attribute identifiers $a \in \mathcal{A}$ to values $v \in \mathcal{V}$ where $\mathcal{A} \cap \mathcal{V} = \emptyset$. A value could be a number, a name (string), a tuple, etc. The scope of a name say $n$, can be restricted by using the restriction operator $\nu n$. For instance, the name $n$ in $C_1 \| \nu n C_2$ is only visible within component $C_2$. The visibility of attribute values can be restricted while the visibility of attribute identifiers is instead never limited. The attribute identifiers represent domain concepts and it is assumed that each component in a system is always aware of them[3]

**Running example (step 1/6):** The robotics scenario can be modeled in *AbC* as follows:

$$Robot_1 \| \ldots \| Robot_n$$

Each robot is modeled as an *AbC* component ($Robot_i$) of the following form ($\Gamma_i{:}P_R$). These components execute in parallel and interact to achieve a specific goal. The attribute *environment* $\Gamma_i$ specifies a set of attributes for each robot. For instance, the attribute "*role*" can take different values like "*explorer*", "*helper*", or "*rescuer*" according to the current state of the robot.    □

---

[3] In the paper, we shall however occasionally use the term "attribute" instead of "attribute identifier".

A *process* is either the *inactive* process $\mathbf{0}$, or a process modeling *action-prefixing* $\bullet.P$ (where "$\bullet$" is replaced with an action), *attribute update* $[\tilde{a} := \tilde{E}]P$, *context awareness* $\langle \Pi \rangle P$, *nodeterministic choice* between two processes $P_1 + P_2$, *parallel composition* of two processes $P_1|P_2$, or *recursive behaviour* $K$ (it is assumed that each process has a unique process definition $K \triangleq P$).

The attribute update construct in $[\tilde{a} := \tilde{E}]P$ sets the value of each attribute in the sequence $\tilde{a}$ to the evaluation of the corresponding expression in the sequence $\tilde{E}$. The awareness construct in $\langle \Pi \rangle P$ is used to test awareness data about a component status or its environment. This construct blocks the execution of process $P$ until the predicate $\Pi$ becomes true. The parallel operator "|" models the interleaving between co-located (i.e., residing within the same component). In what follows, we shall use the notation $[\![\Pi]\!]_\Gamma$ (resp. $[\![E]\!]_\Gamma$) to indicate the evaluation of a predicate $\Pi$ (resp. an expression $E$) under the attribute environment $\Gamma$.

It should be noted that the evaluation of a predicate consists of replacing variable references with their values and returning the result.

**Running example (step 2/6):** We will consider a process $P_R$ running on a robot of the following form:

$$P_R \triangleq (\langle \Pi \rangle a_1.P_1 + a_2.P_2)|P_3$$

The behavior of $P_R$ is a parallel composition of two subprocesses where the one on the left-hand side of "|" can either perform $a_1$ and continue as $P_1$ (if the evaluation of $\Pi$ under the attribute environment semantically equals to true) or perform $a_2$ and continue as $P_2$.    □

In $AbC$ there are two kinds of *communication actions*:

- the attribute-based input $\Pi(\tilde{x})$ which binds to sequence $\tilde{x}$ the corresponding values received from components whose attributes satisfy the predicate $\Pi$;
- the attribute-based output $(\tilde{E})@\Pi$ which evaluates the sequence of expressions $\tilde{E}$ under the attribute environment $\Gamma$ and then sends the result to the components whose attributes satisfy the predicate $\Pi$.

A *predicate* $\Pi$ is either a binary operator $\bowtie$ between two values or a propositional combination of predicates. Predicate $\mathtt{tt}$ is satisfied by all components and is used when modeling broadcast while $\mathtt{ff}$ is not satisfied by any component and is used when modeling silent moves. The satisfaction relation $\models$ of predicates is presented in Table 2. In the rest of this paper, we shall use the relation $\doteq$ to denote a semantic equivalence for predicates as defined below.

**Definition 1 (Predicates Equivalence).** *Two predicates are semantically equivalent, written $\Pi_1 \doteq \Pi_2$, iff for every environment $\Gamma$, it holds that:*

$$\Gamma \models \Pi_1 \text{ iff } \Gamma \models \Pi_2$$

An *expression* $E$ is either a constant value $v \in \mathcal{V}$, or a variable $x$, or an attribute identifier $a$, or a reference to a local attribute value $\mathtt{this}.a$. The properties of *self-awareness* and *context-awareness* that are typical for CAS are guaranteed in *AbC* by referring to the values of local attributes via a special name `this`. (i.e., `this`.*a*). These values represent either the current status of a component (i.e., *self-awareness*) or the external environment as perceived by the component (i.e., *context-awareness*). Expressions within predicates contain also variable names, so predicates can check whether the values that are sent to a specific component do satisfy specific conditions. This permits a sort of pattern-matching. For instance, component $\Gamma{:}(x > 2)(x, y)$ receives a sequence of values "$x, y$" from another component only if the value $x$ is greater than 2.

We assume that our processes are *closed* (i.e., without free process variables), and that free names can be used whenever needed. The constructs $\nu x$ and $\Pi(\tilde{x})$ act as binders for names (i.e., in $\nu x C$ and $\Pi(\tilde{x}).P$, $x$ and $\tilde{x}$ are bound in $C$ and $P$, respectively). We use the notation $bn(P)$ to denote the set of bound names of $P$. The free names of $P$ are those that do not occur in the scope of any binder and are denoted by $fn(P)$. The set of names of $P$ is denoted by $n(P)$. The notions of bound and free names are applied in the same way to components, but free names also include all attribute values that do not occur in the scope of any binder.

**Running example (step 3/6):** By specifying the predicate $\Pi$ and the actions $a_1$ and $a_2$, the process $P_R$ becomes:

$$P_R \triangleq (\langle\texttt{this}.victimPerceived = \texttt{tt}\rangle\,[\texttt{this}.state := stop, \dots]()@\texttt{ff}.P_1 \ + $$
$$(\texttt{this}.id,\ qry,\ \texttt{this}.role)@(role = rescuer \vee role = helping).P_2\ )\mid P_3$$

The process on the left-hand side of "|" , models the situation in which either the robot recognizes the presence of a victim and updates its "*state*" to "*stop*", which triggers it to halt, and the process continues as $P_1$, or it sends a query to ask for information about the position of the victim. This query contains the robot identity "`this`.*id*", a special name "*qry*" to indicate the request type, and the current role of the robot "`this`.*role*". The attributes "*id*" and "*role*" are the only exposed attributes for interaction. Sending on a false predicate "()@ff" models a silent move and the "…" denotes other possible attribute updates.   □

## 3   AbC Operational Semantics

The operational semantics of *AbC* is based on two relations. The transition relation $\longmapsto$ that describes the behaviour of single components and the transition relation $\longrightarrow$ that relies on the former relation and describes systems behaviors.

### 3.1   Operational semantics of component

We use the transition relation $\longmapsto \subseteq Comp \times CLAB \times Comp$ to define the local behavior of a component where $Comp$ denotes a component and $CLAB$ is the set of transition labels $\alpha$ generated by the following grammar:

$$
\begin{array}{ll}
\textbf{Brd} \dfrac{[\![\tilde{E}]\!]_\Gamma = \tilde{v} \quad [\![\Pi_1]\!]_\Gamma = \Pi}{\Gamma : (\tilde{E})@\Pi_1.P \xrightarrow{\overline{\Pi}\tilde{v}} \Gamma : P}
&
\textbf{Rcv} \dfrac{[\![\Pi[\tilde{v}/\tilde{x}]]\!]_\Gamma \simeq \mathrm{tt} \quad \Gamma \models \Pi'}{\Gamma : \Pi(\tilde{x}).P \xrightarrow{\Pi'(\tilde{v})} \Gamma : P[\tilde{v}/\tilde{x}]}
\\[3ex]
\textbf{Upd} \dfrac{[\![\tilde{E}]\!]_\Gamma = \tilde{v} \quad \Gamma[\tilde{a} \mapsto \tilde{v}] : P \xrightarrow{\lambda} \Gamma[\tilde{a} \mapsto \tilde{v}] : P'}{\Gamma : [\tilde{a} := \tilde{E}]P \xrightarrow{\lambda} \Gamma[\tilde{a} \mapsto \tilde{v}] : P'}
&
\textbf{Aware} \dfrac{[\![\Pi]\!]_\Gamma \simeq \mathrm{tt} \quad \Gamma : P \xrightarrow{\lambda} \Gamma' : P'}{\Gamma : \langle \Pi \rangle P \xrightarrow{\lambda} \Gamma' : P'}
\\[3ex]
\textbf{Sum} \dfrac{\Gamma : P_1 \xrightarrow{\lambda} \Gamma' : P_1'}{\Gamma : P_1 + P_2 \xrightarrow{\lambda} \Gamma' : P_1'}
&
\textbf{Rec} \dfrac{\Gamma : P \xrightarrow{\alpha} \Gamma' : P' \quad K \triangleq P}{\Gamma : K \xrightarrow{\alpha} \Gamma' : P'}
\\[3ex]
& \textbf{Int} \dfrac{\Gamma : P_1 \xrightarrow{\lambda} \Gamma' : P_1'}{\Gamma : P_1|P_2 \xrightarrow{\lambda} \Gamma' : P_1'|P_2}
\end{array}
$$

**Table 3.** Component semantics

$$
\alpha ::= \lambda \quad | \quad \widetilde{\Pi(\tilde{v})} \qquad\qquad \lambda ::= \nu\tilde{x}\overline{\Pi}\tilde{v} \quad | \quad \Pi(\tilde{v})
$$

The $\lambda$-labels are used to denote $AbC$ output ($\nu\tilde{x}\overline{\Pi}\tilde{v}$) and input actions ($\Pi(\tilde{v})$). The output and input labels contain the sender's predicate that specifies the communication partners $\Pi$, and the transmitted values $\tilde{v}$. An output is called "bound" if its label contains a bound name (i.e., if $\tilde{x} \neq \emptyset$). The $\alpha$-labels include an additional label $\widetilde{\Pi(\tilde{v})}$ to denote the case where a process is not able to receive a message. As it will be shown later in this section, this kind of labels is crucial to appropriately handle dynamic constructs like choice and awareness. Free names in $\alpha$ are specified as follows:

$$
\begin{aligned}
&- \; fn(\nu\tilde{x}\overline{\Pi}(\tilde{v})) = fn(\Pi(\tilde{v}))\backslash\tilde{x} \qquad \text{and} \qquad fn(\Pi(\tilde{v})) = fn(\Pi) \cup \tilde{v} \\
&- \; fn(\widetilde{\Pi(\tilde{v})}) = fn(\Pi) \cup \tilde{v} \qquad \text{and} \qquad n(\nu\tilde{x}\overline{\Pi}(\tilde{v})) = \emptyset \text{ when } \Pi \simeq \mathrm{ff}
\end{aligned}
$$

The free names of a predicate is the set of names occurring in that predicate except for attribute identifiers. Notice that `this`.$a$ is only a reference to the value of the attribute identifier $a$. Only the output label has bound names (i.e., $bn(\nu\tilde{x}\overline{\Pi}\tilde{v} = \tilde{x}$). The transition relation $\longmapsto$ is formally defined in Table 3 and Table 4.

**Component behavior.** The set of rules in Table 3 describes the behavior of a single $AbC$ component. The symmetrical rule for (**Sum**) and (**Int**) are omitted.

Rule (**Brd**) evaluates the sequence of expressions $\tilde{E}$, say to $\tilde{v}$, and the predicate $\Pi_1$ to $\Pi$ after replacing any reference (i.e., `this`.$a$) with its value according to the attribute environment $\Gamma$, and sends this information in the message, afterwards the process evolves to $P$.

Rule (**Rcv**) replaces the free occurrences of the input sequence variables $\tilde{x}$ in the receiving predicate $\Pi$ with the corresponding message values $\tilde{v}$ and evaluates $\Pi$ under the environment $\Gamma$. If the evaluation semantically equals to $\mathrm{tt}$ and the

$$\textbf{FBrd} \quad \Gamma:(\tilde{E})@\Pi.P \xrightarrow{\widehat{\Pi'(\tilde{v})}} \Gamma:(\tilde{E})@\Pi.P \qquad \textbf{FRcv} \quad \frac{[\![\Pi[\tilde{v}/\tilde{x}]]\!]_\Gamma \neq \mathsf{tt} \;\vee\; (\Gamma \not\models \Pi')}{\Gamma:\Pi(\tilde{x}).P \xrightarrow{\widehat{\Pi'(\tilde{v})}} \Gamma:\Pi(\tilde{x}).P}$$

$$\textbf{FUpd} \quad \frac{[\![\tilde{E}]\!]_\Gamma = \tilde{v} \quad \Gamma[\tilde{a} \mapsto \tilde{v}]:P \xrightarrow{\widehat{\Pi(\tilde{w})}} \Gamma[\tilde{a} \mapsto \tilde{v}]:P}{\Gamma:[\tilde{a} := \tilde{E}]P \xrightarrow{\widehat{\Pi(\tilde{w})}} \Gamma:[\tilde{a} := \tilde{E}]P} \qquad \textbf{FZero} \quad \Gamma:0 \xrightarrow{\widehat{\Pi(\tilde{v})}} \Gamma:0$$

$$\textbf{FAware1} \quad \frac{[\![\Pi]\!]_\Gamma \simeq \mathsf{tt} \quad \Gamma:P \xrightarrow{\widehat{\Pi'(\tilde{v})}} \Gamma:P}{\Gamma:\langle\Pi\rangle P \xrightarrow{\widehat{\Pi'(\tilde{v})}} \Gamma:\langle\Pi\rangle P} \qquad \textbf{FAware2} \quad \frac{[\![\Pi]\!]_\Gamma \simeq \mathsf{ff}}{\Gamma:\langle\Pi\rangle P \xrightarrow{\widehat{\Pi'(\tilde{v})}} \Gamma:\langle\Pi\rangle P}$$

$$\textbf{FSum} \quad \frac{\Gamma:P_1 \xrightarrow{\widehat{\Pi(\tilde{v})}} \Gamma:P_1 \quad \Gamma:P_2 \xrightarrow{\widehat{\Pi(\tilde{v})}} \Gamma:P_2}{\Gamma:P_1 + P_2 \xrightarrow{\widehat{\Pi(\tilde{v})}} \Gamma:P_1 + P_2}$$

$$\textbf{FInt} \quad \frac{\Gamma:P_1 \xrightarrow{\widehat{\Pi(\tilde{v})}} \Gamma:P_1 \quad \Gamma:P_2 \xrightarrow{\widehat{\Pi(\tilde{v})}} \Gamma:P_2}{\Gamma:P_1|P_2 \xrightarrow{\widehat{\Pi(\tilde{v})}} \Gamma:P_1|P_2}$$

**Table 4.** Discarding input

receiver environment $\Gamma$ satisfies the sender predicate $\Pi'$, the input action is performed and the substitution $[\tilde{v}/\tilde{x}]$ is applied to the continuation process $P$.

Rule (**Upd**) evaluates the sequence of expressions $\tilde{E}$ under the environment $\Gamma$, apply attribute updates i.e., $\Gamma[\tilde{a} \mapsto \tilde{v}]$ where $\forall a \in \tilde{a}$ and $\forall v \in \tilde{v}$, we have that: $\Gamma[a \mapsto v](a') = \Gamma(a')$ if $a \neq a'$ and $v$ otherwise, and then performs an action with a $\lambda$ label if process $P$ under the updated environment can do so.

Rule (**Aware**) evaluates the predicate $\Pi$ under the environment $\Gamma$. If the evaluation semantically equals to $\mathsf{tt}$, process $\langle\Pi\rangle P$ proceeds by performing an action with a $\lambda$-label and continues as $P'$ if process $P$ can perform the same action.

Rule (**Sum**) and its symmetric version represent the non-deterministic choice between the subprocesses $P_1$ and $P_2$ in the sense that if any of them say $P_1$ performs an action with a $\lambda$-label and becomes $P_1'$ then the overall process continues as $P_1'$.

Rule (**Rec**) and rule (**Int**) are the standard rules for handling process definition and interleaving of the actions of two processes, respectively.

**Running example (step 4/6):** The process $P_R$ running on a robot, say $Robot_1$ with an attribute environment $\Gamma$ and $\Gamma(id) = 1$, apart from the behavior of $P_3$ (not specified here) or the possibility to discard incoming messages, can either update its attributes $\Gamma : P_R \xrightarrow{\widehat{\mathsf{ff}()}} \Gamma[status \mapsto stop, \ldots]:P_1|P_3$ or send a message $\Gamma : P_R \xrightarrow{\overline{(role=rescuer \;\vee\; role=helping)(1, \; qry, \; explorer)}} \Gamma : P_2|P_3$ . $\qquad\qquad \square$

**Discarding input.** The rules for enabling components to discard specific actions are presented in Table 4, label $\overline{\Pi(\tilde{v})}$ is used to indicate discarding actions. These actions will be needed in the next section when we will define systems semantics.

Rule (**FBrd**) states that any sending process discards messages from other processes and stay unchanged. Rule (**FRcv**) states that if one of the receiving requirements is not satisfied then the process will discard the message and stay unchanged.

Rule (**FUpd**) state that process $[\tilde{a} := \tilde{E}]P$ discards a message if process $P$ is able to discard the same message after applying attribute updates. Rule (**FAware1**) states that process $\langle \Pi \rangle P$ discards a message even if $\Pi$ evaluates to (tt) if process $P$ is able to discard the same message. Rule (**FAware2**) states that if $\Pi$ in process $\langle \Pi \rangle P$ evaluates to ff, process $\langle \Pi \rangle P$ will discard any message from other processes.

Rule (**FZero**) states that process $\mathbf{0}$ always discards messages from other processes. Rule (**FSum**) states that process $P_1 + P_2$ discards a message if both its subprocesses $P_1$ and $P_2$ can do so. The role of the discarding label is to keep dynamic constructs like awareness and choice from dissolving after a message refusal. Rule (**FInt**) has a similar meaning of Rule (**FSum**).

**Running example (step 5/6):** Assume that $Robot_1$ with "*explorer*" role is searching for a victim in some arena and that process $P_3$ in our example can only make silent moves. If a process residing in $Robot_2$ with "*charger*" role sends information about a nearby charging station then process $P_R$ that resides in $Robot_1$ can evolve as follows: $\Gamma : P_R \xrightarrow{\widetilde{(role=explorer)(info)}} \Gamma : P_R$ where $\Gamma$ is the attribute environment of $Robot_1$. Process $P_R$ applies rule (**FInt**) and discards the message because both its subprocesses can discard the message. $P_3$ is not ready for receiving messages, so it applies (**FBrd**) and stay unchanged. The choice also discards the message by applying (**FSum**) because both its subprocesses can discard the message. The subprocess on the left-hand side of $+$ is not ready for receiving messages, so it applies (**FAware2**) and stay unchanged. The subprocess on the right-hand side of $+$ applies (**FBrd**) and discards the message, because it is not ready for receiving messages.                                                    □

### 3.2   Operational semantics of system

*AbC* system describes the global behavior of a component and the underlying communication between different components. We use the transition relation $\longrightarrow \subseteq Comp \times SLAB \times Comp$ to define the behavior of a system where $Comp$ denotes a component and $SLAB$ is the set of transition labels $\gamma$ which are generated by the following grammar:

$$\gamma \; ::= \nu \tilde{x} \overline{\Pi}(\tilde{v}) \quad | \quad \Pi(\tilde{v}) \quad | \quad \tau$$

The $\gamma$-labels extend $\lambda$ with $\tau$ to denote silent moves (i.e., send on a false predicate $()@\text{ff}$). The $\tau$-label has no free or bound names. The definition of the transition relation $\longrightarrow$ depends on the definition of the relation $\longmapsto$ in the previous section in the sense that the effect of local behavior is lifted to the

$$\textbf{Comp}\ \dfrac{\Gamma:P\overset{\lambda}{\longmapsto}\Gamma':P'}{\Gamma:P\overset{\lambda}{\longrightarrow}\Gamma':P'}\qquad\textbf{C-Fail}\ \dfrac{\Gamma:P\overset{\widetilde{\Pi(\tilde v)}}{\longmapsto}\Gamma:P}{\Gamma:P\overset{\Pi(\tilde v)}{\longrightarrow}\Gamma:P}\qquad\textbf{Rep}\ \dfrac{C\overset{\gamma}{\longrightarrow}C'}{!C\overset{\gamma}{\longrightarrow}C'\|!C}$$

$$\tau\textbf{-Int}\ \dfrac{C_1\overset{\nu\tilde x\overline{\Pi}\tilde v}{\longrightarrow}C_1'\quad\Pi\simeq\text{ff}}{C_1\|C_2\overset{\tau}{\longrightarrow}C_1'\|C_2}\qquad\qquad\textbf{Res}\ \dfrac{C[y/x]\overset{\gamma}{\longrightarrow}C'\quad y\notin n(\gamma)\wedge y\notin fn(C)\backslash\{x\}}{\nu xC\overset{\gamma}{\longrightarrow}\nu yC'}$$

$$\textbf{Sync}\ \dfrac{C_1\overset{\Pi(\tilde v)}{\longrightarrow}C_1'\quad C_2\overset{\Pi(\tilde v)}{\longrightarrow}C_2'}{C_1\parallel C_2\overset{\Pi(\tilde v)}{\longrightarrow}C_1'\parallel C_2'}\qquad\textbf{Com}\ \dfrac{C_1\overset{\nu\tilde x\overline{\Pi}\tilde v}{\longrightarrow}C_1'\quad C_2\overset{\Pi(\tilde v)}{\longrightarrow}C_2'\quad\begin{array}{c}\Pi\neq\text{ff}\\\tilde x\cap fn(C_2)=\emptyset\end{array}}{C_1\parallel C_2\overset{\nu\tilde x\overline{\Pi}\tilde v}{\longrightarrow}C_1'\parallel C_2'}$$

$$\textbf{Hide1}\ \dfrac{C\overset{\nu\tilde x\overline{\Pi}\tilde v}{\longrightarrow}C'\quad\begin{array}{c}(\Pi\blacktriangleright y)\simeq\text{ff}\\y\in n(\Pi)\end{array}}{\nu yC\overset{\nu\tilde x\overline{\Pi}\tilde v}{\longrightarrow}\nu y\nu\tilde xC'}\qquad\qquad\textbf{Hide2}\ \dfrac{C\overset{\nu\tilde x\overline{\Pi}\tilde v}{\longrightarrow}C'\quad\begin{array}{c}(\Pi\blacktriangleright y)\neq\text{ff}\\y\in n(\Pi)\end{array}}{\nu yC\overset{\nu\tilde x\overline{\Pi\blacktriangleright y}\tilde v}{\longrightarrow}\nu yC'}$$

$$\textbf{Open}\ \dfrac{C[y/x]\overset{\overline{\Pi}\tilde v}{\longrightarrow}C'\quad\Pi\neq\text{ff}\quad y\in\tilde v\backslash n(\Pi)\wedge y\notin fn(C)\backslash\{x\}}{\nu xC\overset{\nu y\overline{\Pi}\tilde v}{\longrightarrow}C'}$$

**Table 5.** System semantics

global one. The transition relation $\longrightarrow$ is formally defined in Table 5; there the symmetric rules for $\tau$-**Int** and **Com** are omitted.

Rule (**Comp**) states that the relations $\longmapsto$ and $\longrightarrow$ coincide when performing either an input or output actions. Rule (**C-Fail**) states that any component $\Gamma:P$ can discard a message and stay unchanged if its local process is willing to do so. Rule (**Rep**) is standard for replication. Rule ($\tau$-**Int**) and its symmetric rule model the interleaving between components $C_1$ and $C_2$ when performing an internal action (i.e., sending with a false predicate).

Rule (**Res**) states that component $\nu xC$ with a restricted name $x$ can still perform an action with a $\gamma$-label as long as $x$ does not occur in the names of the label and component $C$ can perform the same action. If necessary, we allow renaming with conditions that ensure avoiding name clashing.

Rule (**Sync**) states that two parallel components $C_1$ and $C_2$ can synchronize while performing an input action. This means that the same message is received by both $C_1$ and $C_2$. Rule (**Com**) states that two parallel components $C_1$ and $C_2$ can communicate if $C_1$ can send a message with a predicate that is different from ff and $C_2$ can possibly receive that message.

| | | |
|---|---|---|
| $\text{tt}\blacktriangleright x$ | $=$ | $\text{tt}$ |
| $\text{ff}\blacktriangleright x$ | $=$ | $\text{ff}$ |
| $(a=m)\blacktriangleright x$ | $=$ | $\begin{cases}\text{ff} & \text{if } x=m\\a=m & \text{otherwise}\end{cases}$ |
| $(\Pi_1\wedge\Pi_2)\blacktriangleright x$ | $=$ | $\Pi_1\blacktriangleright x\ \wedge\ \Pi_2\blacktriangleright x$ |
| $(\Pi_1\vee\Pi_2)\blacktriangleright x$ | $=$ | $\Pi_1\blacktriangleright x\ \vee\ \Pi_2\blacktriangleright x$ |
| $(\neg\Pi)\blacktriangleright x$ | $=$ | $\neg(\Pi\blacktriangleright x)$ |

**Table 6.** Predicate restriction $\blacktriangleright x$

Rules (**Hide1**) and (**Hide2**) are unique to $AbC$ and introduce a new concept that we call predicate restriction "$\blacktriangleright x$" as reported in Table 6. In process calculi where broadcasting is the basic primitive for communication like CSP [12] and $b\pi$-calculus [19], broadcasting on a private channel is equal to performing an internal action and no other process can observe the broadcast except the one that performed it.

For example in $b\pi$-calculus, if we let

$P = \nu a(P_1 \| P_2) \| P_3$ where $P_1 = \bar{a}v.Q$,  $P_2 = a(x).R$, and   $P_3 = b(x)$

then if $P_1$ broadcasts on $a$ we would have that only $P_2$ can observe it since $P_2$ is within the scope of the restriction. $P_3$ and other processes only observe an internal action, so $P \xrightarrow{\tau} \nu a(Q\|R[v/x]) \| b(x)$.

This idea is generalized in $AbC$ to what we call predicate restriction "$\bullet\blacktriangleright x$" in the sense that we either hide a part or the whole predicate using the predicate restriction operator "$\bullet\blacktriangleright x$" where $x$ is a restricted name and the "$\bullet$" is replaced with a predicate. If the predicate restriction operator returns ff then we get the usual hiding operator like in CSP and $b\pi$-calculus because the resulting label is not exposed according to ($\tau$-**Int**) rule (i.e., sending with a false predicate).

If the predicate restriction operator returns something different from ff then the message is exposed with a smaller predicate and the restricted name remains private. Note that any private name in the message values (i.e., $\tilde{x}$) remains private if $(\Pi \blacktriangleright y) \mathrel{\hat{=}} \mathsf{ff}$ as in rule (**Hide1**) otherwise it is not private anymore as in rule (**Hide2**). In other words, messages are sent on a channel that is partially exposed.

We would like to stress that the predicate restriction operator, that filters the exposure of the communication predicate either partially or completely, is very useful when modelling user-network interaction. The user observes the network as a single node and interacts with it through a public channel and is not aware of how the messages are propagated through the network. Networks propagate messages between their nodes through private channels while exposing messages to users through public channels. For instance, if a network sends a message with the predicate ($keyword = \mathtt{this}.topic \;\vee\; capability = fwd$) where the name "$fwd$" is restricted then the message is exposed to the user at every node with forwarding capability in the network with this predicate ($keyword = \mathtt{this}.topic$). Network nodes observe the whole predicate but they receive the message only because they satisfy the other part of the predicate (i.e., ($capability = fwd$)). In the following Lemma, we prove that the satisfaction of a restricted predicate $\Pi \blacktriangleright x$ by an attribute environment $\Gamma$ does not depend on the name $x$ that is occurring in $\Gamma$.

**Lemma 1.** $\Gamma \models \Pi \blacktriangleright x$   iff   $\forall v.\ \Gamma[v/x] \models \Pi \blacktriangleright x$   for any environment $\Gamma$, predicate $\Pi$, and name $x$.

Rule (**Open**) states that a component has the ability to communicate a private name to other components. This rule is different from the one in $\pi$-calculus in the sense that $AbC$ represents multiparty settings. This implies that the scope of the private name $x$ is not expanded to include a group of other components but rather the scope is dissolved. In other words, when a private name is communicated in $AbC$ then the name is not private anymore. Note that, a component that is sending on a false predicate (i.e., $\Pi \mathrel{\hat{=}} \mathsf{ff}$) cannot open the scope.

**Running example (step 6/6):** If we further specify the subprocess $P_1$ in the process running on the robots, $P_R$ becomes:

$$P_R \triangleq (\langle \texttt{this}.victimPerceived = \texttt{tt} \rangle \, [\ldots, \texttt{this}.role := rescuer]()@\texttt{ff}.$$
$$(y = qry \wedge z = explorer)(x,\ y,\ z).P_1' \quad +$$
$$(\texttt{this}.id,\ qry,\ \texttt{this}.role)@(role = rescuer \vee role = helping).P_2 \ ) \mid P_3$$

Basically, the robot changes its role to "*rescuer*" after recognizing the victim and waits for queries from nearby explorers. Once a query is received, the local process continues as $P_1'$.

Let us assume that the role of $Robot_1$ is "*rescuer*" and $Robot_2$ is "*explorer*". $Robot_2$ can send a query to nearby rescuing or helping robots (i.e., $Robot_1$) by using rule (**Comp**) and generate this transition:

$$Robot_2 \xrightarrow{\overline{(role=rescuer \vee role=helping)}(2,\ qry,\ explorer)} \Gamma_2 : (P_2 | P_3)$$

On the other hand, $Robot_1$ can receive this query by using rule (**Comp**) and generate this transition:

$$Robot_1 \xrightarrow{(role=rescuer \vee role=helping)(2,\ qry,\ explorer)}$$
$$\Gamma_1 : (P_1'[2/x,\ qry/y,\ explorer/z] | P_3)$$

Other robots which are not addressed by communication discard the message by applying rule (**C-Fail**). Now the overall system evolves by applying rule (**Com**) as follows:

$$S \xrightarrow{\overline{(role=rescuer \vee role=helping)}(2,\ qry,\ explorer)}$$
$$\Gamma_1 : (P_1'[2/x,\ qry/y,\ explorer/z] | P_3) \parallel \Gamma_2 : (P_2 | P_3) \parallel \Gamma_3 : P_{R_3} \parallel \ldots \parallel \Gamma_n : P_{R_n}$$
$$\square$$

## 4 Expressiveness of AbC Calculus

In this section, we provide evidence of the expressive power of $AbC$ by presenting a complete model for the swarm robotics scenario and by showing how other communication models can be easily rendered with $AbC$. We provide evidence that $AbC$ can be used to naturally model systems featuring collaboration, adaptation, and reconfiguration and advocate the use of attribute-based communication as a unifying framework to encompass different communication models.

### 4.1 A swarm robotics model in *AbC*

The swarm robotics model exploits the fact that a process running on a robot can either read the values of some attributes that are provided by its sensors or read and update the other attributes in its attribute environment. Reading the values of the attributes controlled by sensors either provides information about the robot environment or information about the current status of the robot. We could say that in the former case the model formalises *context-awareness* while in the latter case it formalizes *self-awareness*. For instance, when reading the value of the *collision* attribute in the attribute environment $\Gamma(collision) = \texttt{tt}$ the robot becomes aware that a collision with a wall in the arena is imminent and this triggers an adaptation mechanism to change the robot direction. On the other hand, reading the value of the *batteryLevel* attribute $\Gamma(batteryLevel) = 15\%$ makes the robot aware that its battery level is critical (i.e., $< 20\%$) and this triggers an adaptation mechanism to halt the movement and to take the robot into the power saving mode.

We assume that each robot has a unique identity ($id$) and since the robot acquires information about its environment or its own status by reading the values provided by sensors, no additional assumption about its initial state is needed. It is worth mentioning that sensors and actuators are not modelled by $AbC$ because they represent the robot internal infrastructure while $AbC$ model represents the programmable behaviour of the robot (i.e., its running code).

The robotics scenario is modelled as a set of parallel $AbC$ components, each of which represents a robot $(Robot_1 \| \ldots \| Robot_n)$ and each robot has the following form $(\Gamma_i : P_R)$. The behaviour of a single robot is modelled in the following $AbC$ process $P_R$:

$$P_R \ \triangleq (Rescuer \ \ + \ \ Explorer) | \ RandWalk \ | \ IsMoving$$

The robot follows a random walk in exploring the disaster arena. The robot can become a "rescuer" when he becomes aware of the presence of a victim by locally reading the value of an attribute controlled by its sensors or remain an "explorer" and keep sending queries for information about the victim from nearby robots whose role is either "*rescuer*" or "*helper*".

If sensors recognise the presence of a victim and the value of "*victimPerceived*" becomes "tt", the robot updates its "*state*" to "*stop*" (which triggers an actuation signal to halt the actuators and stop movement), computes the victim position and the number of the required robots to rescue the victim and stores them in the attributes "*vPosition*" and "*count*" respectively, changes its role to "*rescuer*", and waits for queries from nearby explorers. Once a message from an explorer is received, the robot sends back the victim information to the requesting robot addressing it by its identity "*id*" and the collective (i.e., the swarm) starts forming in preparation for the rescuing procedure.

$$Rescuer \triangleq \langle \texttt{this}.victimPerceived = \texttt{tt} \rangle [\texttt{this}.state := stop, \ \texttt{this}.count := 3,$$
$$\texttt{this}.vPosition := <3, 4>, \ \texttt{this}.role := rescuer]()@\texttt{ff}.$$
$$(y = qry \land z = explorer)(x, \ y, \ z).$$
$$(\texttt{this}.vPosition, \ \texttt{this}.count, \ ack, \ \texttt{this}.role)@(id = x)$$

On the other hand, if the victim is still not perceived, the robot continuously sends queries for information about the victim to the nearby robots whose role is either "rescuer" or "*helper*". The query message contains the robot identity "*this.id*" , a special name "*qry*" to indicate the request type, and the current role of the robot "*this.role*". If an acknowledgement arrives containing victim's information, the robot changes its role to "*helper*" and starts the helping procedure.

$$Explorer \triangleq (\texttt{this}.id,\ qry,\ \texttt{this}.role)@(role = rescuer \lor role = helper).$$
$$(((z = rescuer \lor z = helper) \land x = ack)(vpos,\ c,\ x,\ z).$$
$$[\texttt{this}.role := helper]()@\texttt{ff}.Helper + Rescuer + Explorer)$$

*Remark 1.* The interaction between an explorer robot currently running and a rescuer robot that is waiting for a request from nearby explorers suggests a possible way of modelling binary communication like in $\pi$-calculus [19]. Rendezvous can be modelled in a similar way by defining an attribute to count the number of needed acknowledgments to signal synchronisation.

The "*Helper*" process defined below is triggered by receipt of the victim information from the rescuer-collective as mentioned above.

$$Helper \triangleq [\texttt{this}.vPosition := vpos,\ \texttt{this}.target := vpos]()@\texttt{ff}.$$
$$(\langle \texttt{this}.position = \texttt{this}.target \rangle [\texttt{this}.role := rescuer]()@\texttt{ff}$$
$$|\ \langle c > 1 \rangle (y = qry \land z = explorer)(x,\ y,\ z).$$
$$(\texttt{this}.vPosition,\ c - 1,\ ack,\ \texttt{this}.role)@(id = x))$$

The helping robot stores the victim position in the attribute "*vPosition*" and updates its target to be the victim position. This triggers the actuators to move to the specified location. The robot moves towards the victim but at the same time is willing to respond to other robots queries, in case more than one robot is needed for the rescuing procedure. Once the robot reaches the victim (i.e., its position coincides with the victim position), the robot changes its role to "*rescuer*" and joins the rescuer-collective.

The "*RandWalk*" process is defined below. This process computes a random direction to be followed by the robot. Once a collision is detected by the proximity sensor, a new random direction is calculated.

$$RandWalk \triangleq [\texttt{this}.direction := 2\pi rand()]()@\texttt{ff}.$$
$$\langle \texttt{this}.collision = \texttt{tt} \rangle RandWalk$$

Finally, process "*IsMoving*" captures the status of the battery level in a robot at any time. Once the battery level drops into a critical level (i.e., less than 20%), the robot changes its status to "*stop*" which results in halting the actuators and the robot enters the power saving mode. The robot stays in this mode until it is recharged to at least 90% and then it starts moving again.

$$IsMoving \triangleq \langle \texttt{this}.state = move \land \neg(\texttt{this}.batteryLevel > 20\%) \rangle$$
$$[\texttt{this}.state := stop]()@\texttt{ff}.\langle \texttt{this}.batteryLevel \geq 90\% \rangle$$
$$[\texttt{this}.state := move]()@\texttt{ff}.IsMoving$$

For simplifying the presentation, in this scenario we are not modelling the charging task and assume that this task is accomplished according to some predefined procedure. It is worth mentioning that if more victims are found in the arena, different rescuer-collectives will be spontaneously formed to rescue them. To avoid forming multiple collectives for the same victim, we assume that sensors only detect isolated victims. Light-based message communication [22] between robots can be used. Thus once a robot has reached a victim, it signals with a specific color light to other robots not to discover the victim next to it [23]. Since we do not model the failure recovery in this scenario, we assume that all robots are fault-tolerant and they cannot fail. For more details, a runtime environment for supporting the linguistic primitives of $AbC$ can be found at the following website http://lazkany.github.io/AbC/. There we provide also a short tutorial to provide some intuition about how to use these primitives for programming.

### 4.2   Encoding channel-based interaction

The interaction primitives in $AbC$ are purely based on attributes. In contrast to other process calculi where senders and receivers have to agree on an explicit channel or name, $AbC$ relies on the satisfaction of predicates over attributes for deriving the interaction.

Attribute values in $AbC$ can be modified by means of internal actions. Changing attribute values permits opportunistic interaction between components in the sense that an attribute update might provide new opportunities of interaction. This is because the selection of interaction partners depends on predicates over the attributes they expose. Changing the values of these attributes implies changing the set of possible partners and this is why modelling adaptivity in $AbC$ is quite natural. Offering this possibility is difficult in channel-based process calculi. Indeed, we would like to argue that finding a compositional encoding in channel-based process calculi for the following simple $AbC$ system is very difficult, if not impossible :

$$\Gamma_1 : (msg, \texttt{this}.b)@(\texttt{tt}) \| \ \Gamma_2 : ([\texttt{this}.a := 5]()@\texttt{ff}.P \mid (y \leq \texttt{this}.a)(x, y).Q)$$

If we assume that initially $\Gamma_1(b) = 3$ and $\Gamma_2(a) = 2$, we have that changing the value of the local attribute $a$ to "5" by the left-hand side process in the second component gives it an opportunity of receiving the message "$msg$" from the process residing in the first component.

Looking from the opposite perspective one might ask whether it is possible to model channel based message passing in $AbC$. Indeed, a feature that is not present in $AbC$ is the possibility of specifying a single name/channel where the exchange happens instantaneously, i.e., the possibility of relying on a channel that appears at the time of interaction and disappears afterwards. Attributes are always available in the attribute environment and cannot disappear when one would like them to do so. However, this is not a problem, since it is possible to exploit the fact that $AbC$ predicates can check the message values. Thus, we can add the name of the channel where the exchange happens as a value in the message. The receiver is left with the responsibility to check the compatibility of that value with its receiving channel.

To show the correctness of this encoding, we choose $b\pi$-calculus [8] as a representative for channel-based process calculi. The $b\pi$-calculus is a good choice because it uses broadcast instead of binary communication as a basic primitive for interaction which makes it a sort of variant of value-passing CBS [25]. Furthermore, channels in $b\pi$-calculus can be communicated like in $\pi$-calculus [19] which is considered as one of the richest paradigms introduced for concurrency so far. Based on a separation results presented in [9], it has been proved that $b\pi$-calculus and $\pi$-calculus are incomparable in the sense that there does not exist any uniform, parallel-preserving translation from $b\pi$-calculus into $\pi$-calculus up to any "reasonable" equivalence. On the other hand, in $\pi$-calculus a process can non-deterministically choose the communication partner while in $b\pi$-calculus cannot.

Proving the existence of a uniform and parallel-preserving encoding of $b\pi$-calculus into $AbC$ up to some reasonable

$$(\text{Component Level})$$
$$\{|G|\}_c \triangleq \emptyset : \{|G|\}_p \qquad \{|P\|Q|\}_c \triangleq \{|P|\}_c \parallel \{|Q|\}_c$$
$$\{|\nu x P|\}_c \triangleq \nu x \{|P|\}_c$$
$$(\text{Process Level})$$
$$\{|\texttt{nil}|\}_p \triangleq 0 \qquad \{|\tau.G|\}_p \triangleq ()@\texttt{ff}.\{|G|\}_p$$
$$\{|a(\tilde{x}).G|\}_p \triangleq \Pi(y,\tilde{x}).\{|G|\}_p$$
$$\quad \texttt{with} \quad \Pi = (y=a) \quad \texttt{and} \quad y \notin n(\{|G|\}_p)$$
$$\{|\bar{a}\tilde{x}.G|\}_p \triangleq (a,\tilde{x})@(a=a).\{|G|\}_p$$
$$\{|(rec\,A\langle\tilde{x}\rangle).G\rangle|\}_p \triangleq (A(\tilde{x}) \triangleq \{|G|\}_p)$$
$$\quad\quad\quad\quad\quad \texttt{where} \quad fn(\{|G|\}_p) \subseteq \{\tilde{x}\}$$
$$\{|G_1+G_2|\}_p \triangleq \{|G_1|\}_p + \{|G_2|\}_p$$

**Table 7.** Encoding b$\pi$-calculus into $AbC$

equivalence ensures at least the same separation results between $AbC$ and $\pi$-calculus.

We consider two level syntax of $b\pi$-calculus (i.e., only static contexts [18] are considered) as shown below.

$$P ::= G \mid P_1 \| P_2 \mid \nu x P$$

$$G ::= \texttt{nil} \mid a(\tilde{x}).G \mid \bar{a}\tilde{x}.G \mid G_1 + G_2 \mid (rec\,A\langle\tilde{x}\rangle.G)\langle\tilde{y}\rangle$$

Dealing with the one level $b\pi$-syntax would not add any difficulty concerning channel encoding; only the encoding of parallel composition and name restriction occurring under a prefix or a choice would be slightly more difficult. As reported in Table 7, the encoding of a $b\pi$-calculus process $P$ is rendered as an $AbC$ component $\{|P|\}_c$ with $\Gamma = \emptyset$. The channel is rendered as the first element in the sequence of values. For instance, in the output action $(a,\tilde{x})@(a=a)$, $a$ represents the interaction channel, so the input action $(y=a)(y,\tilde{x})$ will always check the first element of the received values to decide whether to accept or discard the message. Notice that the predicate $(a=a)$ is satisfied by any $\Gamma$, however including the channel name in the predicate is crucial to encode name restriction correctly.

The formal definition which specifies what properties are preserved by this encoding and a proof sketch for the correctness of the encoding up to a specific behavioral equivalence will be presented in Section 5.3.

### 4.3   Encoding interaction patterns

In this section, we provide insights on how the concept of *attribute-based communication* can be exploited to provide a general unifying framework encompassing different interaction patterns tailored for multiway interactions. We show how group-based [1, 4, 13] and publish/subscribe-based [3, 10] interaction patterns can be naturally rendered in $AbC$. Since these interaction patterns do not have formal descriptions, we proceed by relying on examples.

We start with group-based interaction patterns and show that when modelling a group name as an attribute in $AbC$, the constructs for joining or leaving a given group can be modelled as attribute updates, like in the following example:

$$\Gamma_1 : (msg, \text{this}.group)@(group = a) \parallel$$
$$\Gamma_2 : ((y = b)(x, \ y) \mid [\text{this}.group := c]()\text{@ff}) \parallel \ldots$$
$$\parallel \ \Gamma_7 : ((y = b)(x, \ y) \mid [\text{this}.group := a]()\text{@ff})$$

We assume that initially we have $\Gamma_1(group) = b$, $\Gamma_2(group) = a$, and $\Gamma_7(group) = c$. Component 1 wants to send the message "$msg$" to group "$a$". Only Component 2 is allowed to receive it as it is the only member of group "$a$". Component 2 can leave group "$a$" and join "$c$" by performing an attribute update with a silent move. On the other hand, if Component 7 joined group "$a$" before "$msg$" is emitted then both of Component 2 and Component 7 will receive the message.

It is worth mentioning that a possible encoding of group communication into $b\pi$-calculus has been introduced in [8]. The encoding is relatively complicated and does not guarantee the causal order of message reception. "Locality" is neither a first class construct in $b\pi$-calculus nor in $AbC$. However, "locality" (in this case, the group name) can be naturally modeled as an attribute in $AbC$ while in $b\pi$-calculus, more efforts are needed.

Publish/subscribe interaction patterns can be considered as special cases of the attribute-based ones. For instance, a natural way of modelling the topic-based publish/subscribe model [10] with $AbC$ would be to allow publishers to broadcast messages with "tt" predicates (i.e., satisfied by all) and permit only the subscribers to check the compatibility of the exposed publishers attributes with their subscriptions. Consider the following example:

$$\Gamma_1 : (msg, \text{this}.topic)@(\text{tt}) \parallel \Gamma_2 : (y = \text{this}.subscription)(x, \ y) \parallel$$
$$\ldots \parallel \Gamma_n : (y = \text{this}.subscription)(x, \ y)$$

The publisher broadcasts the message "$msg$" tagged with a specific topic for all possible subscribers (the predicate "tt" is satisfied by all), subscribers receive the message if the topic matches their subscription.

## 5    Behavioral Theory for AbC

In this section, we define a behavioral theory for $AbC$. We start by introducing a reduction barbed congruence, then we present an equivalent definition of a labeled bisimulation. At the end of the section, we sketch the proof of the correctness of the encoding of Section 4.2 up to strong reduction barbed congruence.

### 5.1   Reduction barbed congruence

In the behavioral theory, two terms are considered as equivalent if they cannot be distinguished by any external observer (i.e., they have the same observable behavior). For instance, in $\pi$-calculus both message transmission and reception are considered to be observable. However, this is not the case in $AbC$ because sending is non-blocking and only message transmission can be observed. It is important to notice that the transition $C \xrightarrow{\overline{\Pi}\tilde{v}} C'$ does not necessarily means that $C$ has performed an input action but rather it means that $C$ *might* have performed an input action. Indeed, this transition might happen due to the application of one of two different rules in Table 5, namely (**Comp**) which guarantees reception and (**C-Fail**) which models non-reception. Hence, input actions cannot be observed by an external observer and only output actions are observable in $AbC$. By following Milner and Sangiorgi [20], we use the term "barb" as synonymous with observable. In what follows, we shall use the following notations:

- $\Rightarrow$ denotes $\xrightarrow{\tau}{}^*$  where  $\tau = \nu\tilde{x}\overline{\Pi}\tilde{v}$  with  $\Pi \simeq \mathsf{ff}$.
- $\xRightarrow{\gamma}$ denotes $\Rightarrow \xrightarrow{\gamma} \Rightarrow$ if $(\gamma \neq \tau)$.
- $\xRightarrow{\hat{\gamma}}$ denotes $\Rightarrow$ if $(\gamma = \tau)$ and $\xRightarrow{\gamma}$ otherwise.
- $\rightarrowtail$ denotes $\{\xrightarrow{\gamma} \mid \gamma$ is an output or $\gamma = \tau\}$   and   $\rightarrowtail^*$ denotes$(\rightarrowtail)^*$

A context $\mathcal{C}[\bullet]$ is a component term with a hole, denoted by $[\bullet]$ and $AbC$ contexts are generated by the following grammar:

$$\mathcal{C}[\bullet] ::= [\bullet] \quad | \quad [\bullet]\|C \quad | \quad C\|[\bullet] \quad | \quad \nu x[\bullet] \quad | \quad ![\bullet]$$

**Definition 2 (Barb).** *Let* $C{\downarrow_\Pi}^4$ *mean that component $C$ can send a message with a predicate $\Pi'$ (i.e., $C \xrightarrow{\nu\tilde{x}\overline{\Pi'}\tilde{v}}$ where $\Pi' \simeq \Pi$ and $\Pi' \not\simeq \mathsf{ff}$). We write $C \Downarrow_\Pi$ if $C \rightarrowtail^* C' \downarrow_\Pi$.*

**Definition 3 (Barb Preservation).** *$\mathcal{R}$ is barb-preserving iff for every $(C_1, C_2) \in \mathcal{R}$, $C_1{\downarrow_\Pi}$ implies $C_2 \Downarrow_\Pi$*

**Definition 4 (Reduction Closure).** *$\mathcal{R}$ is reduction-closed iff for every $(C_1, C_2) \in \mathcal{R}$, $C_1 \rightarrowtail C_1'$ implies $C_2 \rightarrowtail^* C_2'$ and $(C_1', C_2') \in \mathcal{R}$*

**Definition 5 (Context Closure).** *$\mathcal{R}$ is context-closed iff for every $(C_1, C_2) \in \mathcal{R}$ and for all contexts $\mathcal{C}[\bullet]$, $(\mathcal{C}[c_1], \mathcal{C}[c_2]) \in \mathcal{R}$*

Now, everything is in place to define reduction barbed congruence.

**Definition 6 (Weak Reduction Barbed Congruence).** *A symmetric relation $\mathcal{R}$ over the set of* AbC-*components which is barb-preserving, reduction-closed, and context-closed, is weak reduction barbed congruence.*

Two components are weak barbed congruent, written $C_1 \cong C_2$, if $(C_1, C_2) \in \mathcal{R}$ for some barbed congruent relation $\mathcal{R}$. The strong reduction congruence "$\simeq$" is obtained in a similar way by replacing $\Downarrow$ with $\downarrow$ and $\rightarrowtail^*$ with $\rightarrowtail$ .

---

[4] From now on, we use the predicate $\Pi$ to denote only its meaning, not its syntax.

**Lemma 2.** *if $C_1 \cong C_2$ then*

- $C_1 \Downarrow_\Pi$ *iff* $C_2 \Downarrow_\Pi$
- $C_1 \rightarrow^* C_1'$ *implies* $C_2 \rightarrow^* \cong C_1'$

### 5.2   Bisimulation Proof Methods

In this section, we define an appropriate notion of bisimulation for *AbC* components. We prove that our bisimilarity coincides with reduction barbed congruence, and thus represents a valid tool for proving that two components are reduction barbed congruent.

**Definition 7 (Weak Bisimulation).** *A symmetric binary relation $\mathcal{R}$ over the set of* AbC-*components is a weak bisimulation if for every action $\gamma$, whenever $(C_1, C_2) \in \mathcal{R}$ and $\gamma$ is of the form $\tau$, $\Pi(\tilde{v})$, or ($\nu\tilde{x}\overline{\Pi}\tilde{v}$ with $\Pi \neq$ ff), it holds that:*

$$C_1 \xrightarrow{\gamma} C_1' \text{ implies } C_2 \stackrel{\hat{\gamma}}{\Rightarrow} C_2' \text{ and } (C_1', C_2') \in \mathcal{R}$$

*where every predicate $\Pi$ occurring in $\gamma$ is matched by its semantics meaning in $\hat{\gamma}$. Two components $C_1$ and $C_2$ are weak bisimilar, written $C_1 \approx C_2$ if there exists a weak bisimulation $\mathcal{R}$ relating them. Strong bisimilarity, "$\sim$", is defined in a similar way by replacing $\Rightarrow$ with $\rightarrow$.*

It is easy to prove that $\sim$ and $\approx$ are equivalence relations by relying on the classical arguments of [18]. However, our bisimilarity enjoys a much more interesting property: the closure under any context. So, in the next three lemmas, we prove that our bisimilarity is preserved by parallel composition, name restriction, and replication.

**Lemma 3 ($\sim$ and $\approx$ are preserved by parallel composition).** *Let $C_1$ and $C_2$ be two components such that:*

- $C_1 \sim (resp. \approx) C_2$ *implies* $C_1 \| C \sim (resp. \approx) C_2 \| C$ *for all components $C$.*

**Lemma 4 ($\sim$ and $\approx$ are preserved by name restriction).** *Let $C_1$ and $C_2$ be two components such that:*

- $C_1 \sim (resp. \approx) C_2$ *implies* $\nu x C_1 \sim (resp. \approx) \nu x C_2$ *for all names $x$.*

**Lemma 5 ($\sim$ and $\approx$ are preserved by replication).** *Let $C_1$ and $C_2$ be two components such that:*

- $C_1 \sim (resp. \approx) C_2$ *implies* $!C_1 \sim (resp. \approx) !C_2$.

As an immediate consequence of Lemma 3, Lemma 4, and Lemma 5, we have that $\sim$ and $\approx$ are congruence relations (i.e., closed under any context ). We are now set to show that our bisimilarity represents a proof technique for establishing reduction barbed congruence.

**Theorem 1 (Soundness).** *Let $C_1$ and $C_2$ be two components such that:*

– *$C_1 \sim (resp. \approx) C_2$ implies $C_1 \simeq (resp. \cong) C_2$.*

Finally, we prove that our bisimilarity is more than a proof technique, but rather it represents a complete characterization of the reduction barbed congruence.

**Lemma 6 (Completeness).** *Let $C_1$ and $C_2$ be two components such that:*

– *$C_1 \simeq (resp. \cong) C_2$ implies $C_1 \sim (resp. \approx) C_2$.*

As a direct consequence of Theorem 1 and Lemma 6, we have that bisimilarity and reduction barbed congruence coincide.

**Theorem 2 (Characterization).** *Bisimilarity and reduction barbed congruence coincide.*

### 5.3   Correctness of the encoding

In this section, we provide a sketch of the proof of correctness of the encoding presented in Section 4.2. We begin by listing the properties that we would like our encoding to preserve. Basically, when translating a term from $b\pi$-calculus into $AbC$, we would like the translation: to be compositional in the sense that it is independent from contexts; to be independent from the names of the source term (i.e., name invariance); to preserve the parallel composition (i.e., homomorphic w.r.t. '|'); to be faithful in the sense it preserves the observable behaviour (i.e., barbs) and reflects divergence; to translate output (input) action in $b\pi$-calculus into a corresponding output (input) in $AbC$, and finally the translation should preserve the operational correspondence between the source and target calculus. This includes that the translation should be complete (i.e., every computation of the source term can be mimicked by its translation) and it should be sound (i.e., every computation of a translated term corresponds to some computation of its source term).

**Definition 8 (Divergence).** *$P$ diverges, written $P \Uparrow$, iff $P \rightarrowtail^\omega$ where $\omega$ denotes an infinite number of reductions.*

**Definition 9 (Uniform Encoding).** *An encoding $( \! | \, . \, | \! ) : \mathcal{L}_1 \rightarrow \mathcal{L}_2$ is uniform if it enjoys the following properties:*

1. *(Homomorphic w.r.t. parallel composition):* $( \! | \, P \| Q \, | \! ) \triangleq ( \! | \, P \, | \! ) \| ( \! | \, Q \, | \! )$
2. *(Name invariance):* $( \! | \, P\sigma \, | \! ) \triangleq ( \! | \, P \, | \! )\sigma$, *for any permutation of names $\sigma$.*
3. *(Faithfulness):* $P \Downarrow_1$ iff $( \! | \, P \, | \! ) \Downarrow_2$; $P \Uparrow_1$ iff $( \! | \, P \, | \! ) \Uparrow_2$
4. *Operational correspondence*
   1. *(Operational completeness): if $P \rightarrowtail_1 P'$ then $( \! | \, P \, | \! ) \rightarrowtail^*_2 \simeq_2 ( \! | \, P' \, | \! )$ where $\simeq$ is the strong barbed equivalence of $\mathcal{L}_2$.*
   2. *(Operational soundness): if $( \! | \, P \, | \! ) \rightarrowtail_2 Q$ then there exists a $P'$ such that $P \rightarrowtail^*_1 P'$ and $Q \rightarrowtail^*_2 \simeq_2 ( \! | \, P' \, | \! )$, where $\simeq$ is the strong barbed equivalence of $\mathcal{L}_2$.*

**Lemma 7 (Operational Completeness).** *if* $P \twoheadrightarrow_{b\pi} P'$ *then* $(\!|P|\!)_c \twoheadrightarrow^* \simeq (\!|P'|\!)_c$.

*Proof.* (Sketch) The proof proceeds by induction on the shortest transition of $\twoheadrightarrow_{b\pi}$. We have several cases depending on the structure of the term $P$. We only consider the case of parallel composition when communication happens: $P_1 \| P_2 \xrightarrow{\nu\tilde{y}\tilde{a}\tilde{z}} P_1' \| P_2'$. By applying induction hypotheses on the premises $P_1 \xrightarrow{\nu\tilde{y}\tilde{a}\tilde{z}} P_1'$ and $P_2 \xrightarrow{a(\tilde{z})} P_2'$, we have that $(\!|\, P_1 \,|\!)_c \twoheadrightarrow^* \simeq (\!|\, P_1' \,|\!)_c$ and $(\!|\, P_2 \,|\!)_c \twoheadrightarrow^* \simeq (\!|\, P_2' \,|\!)_c$. We can apply rule (Com).

$$\frac{\emptyset : (\!|P_1|\!)_p \xrightarrow{\nu\tilde{y}\overline{(a=a)}(a,\tilde{z})} \emptyset : (\!|P_1'|\!)_p \quad \emptyset : (\!|P_2|\!)_p \xrightarrow{(a=a)(a,\tilde{z})} \emptyset : (\!|P_2'|\!)_p}{\emptyset : (\!|P_1|\!)_p \ \| \ \emptyset : (\!|P_2|\!)_p \xrightarrow{\nu\tilde{y}\overline{(a=a)}(a,\tilde{z})} \emptyset : (\!|P_1'|\!)_p \ \| \ \emptyset : (\!|P_2'|\!)_p}$$

Now, it is easy to see that: $(\!|P_1'\|P_2'|\!)_c \simeq \emptyset : (\!|P_1'|\!)_p \| \ \emptyset : (\!|P_2'|\!)_p$. Notice that the $b\pi$ term and its encoding have the same observable behavior i.e., $P_1 \| P_2 \downarrow_a$ and $(\!|P_1\|P_2|\!)_c \downarrow_{(a=a)}$. $\qquad\square$

**Lemma 8 (Operational Soundness).** *if* $(\!|P|\!)_c \twoheadrightarrow Q$ *then* $\exists P'$ *such that* $P \twoheadrightarrow_{b\pi} P'$ *and* $Q \twoheadrightarrow^* \simeq (\!|P'|\!)_c$.

The idea that we can mimic each transition of $b\pi$-calculus by exactly one transition in $AbC$ implies that soundness and completeness of the operational correspondence can be even proved in a stronger way as in corollary 1 and 2.

**Corollary 1 (Strong Completeness).** *if* $P \twoheadrightarrow_{b\pi} P'$ *then* $\exists Q$ *such that* $Q \equiv (\!|P'|\!)_c$ *and* $(\!|P|\!)_c \twoheadrightarrow Q$.

**Corollary 2 (Strong Soundness).** *if* $(\!|P|\!)_c \twoheadrightarrow Q$ *then* $Q \equiv (\!|P'|\!)_c$ *and* $P \twoheadrightarrow_{b\pi} P'$

**Theorem 3.** *The encoding* $(\!|\ \bullet\ |\!) : b\pi \to \text{AbC}$ *is uniform.*

*Proof.* Definition 9(1) and 9(2) hold by construction. Definition 9(4) holds by Lemma 7, Lemma 8, Corollary 1, and Corollary 2 respectively. Definition 9(3) holds easily and as a result of the proof of Lemma 7 and the strong formulation of operational correspondence in Corollary 1, and Corollary 2, this encoding preserves the observable behavior and cannot introduce divergence. $\qquad\square$

## 6  Discussion and Related Work

In this section, we discuss the main differences between $AbC$ and its old version presented in [2], then we touch on related works concerning calculi with primitives that model either collective interaction or multiparty interaction.

The old $AbC$ is very basic and has many limitations. In this paper, we have fully redesigned the calculus and added essential features needed to effectively control interactions in an attribute-based framework. The extended version of *AbC* includes: polyadic (i.e., multi-valued) communication, replication, name restriction, multithreading, awareness constructs, and a richer language for defining predicates.

The old *AbC* does not support awareness since its components have no explicit way to read/check their attribute environments and react accordingly. This greatly impacts on expressiveness in that awareness-based applications can hardly be tackled. The old calculus has also problems in modeling adaptation; the impossibility of accessing the attribute environment implies that interaction predicates are static and cannot take into account the changing attributes. For instance, in the robotic scenario in Section 4.1, if a collision with a wall in the arena is detected, the robot cannot adapt by changing its direction because the robot is not aware of its environment. The same applies for its own status e.g., its battery level, so modeling such a scenario in old *AbC* is hardly possible.

In the old calculus, the whole component state is always exposed during interaction in the sense that different messages sent by a component are characterized by the same set of attributes. A simple behavior like $(\bar{a}c + \bar{b}d) \parallel (a(x) + b(x))$ – where action $a$ (resp. $b$) synchronizes only with action $a$ (resp. $b$) – cannot be modeled. This behavior can be simply modeled in *AbC* as follows:

$$\Gamma_1{:}(a,\ c)@(\mathsf{tt}) + (b,\ d)@(\mathsf{tt}) \quad \parallel \quad \Gamma_2{:}(x=a)(x,\ y) + (x=b)(x,\ y)$$

By including the channel names (i.e., $a$ and $b$ ) in the message we are guaranteed that message $c$ will only be received by the process with predicate $(x = a)$ and message $d$ will only be received by the process with predicate $(x = b)$. Clearly, the absence of multi-value passing in the old calculus impacts its expressiveness and makes modeling channel-based communication very hard.

As mentioned in the introduction, *AbC* is inspired by the SCEL language [6, 7] that was designed to support programming of autonomic computing systems [27]. Compared with SCEL, the knowledge representation in *AbC* is abstract and is not designed for detailed reasoning during the model evolution. This reflects the different objectives of SCEL and *AbC*. While SCEL focuses on programming issues, *AbC* concentrates on a minimal set of primitives to study attribute-based communication.

In [26], a survey of formal methods for supporting swarm/collective behavior is presented. The results show that there does not exist a single formalism to support such kind of behavior, but different formalisms can be combined to reach this goal. Due to the simplicity of process calculi, specification can become large and therefore difficult to read and understand. Most process calculi cannot explicitly deal with data or algorithmic issues. They do not support modeling and reasoning about persistent information so adaptive behavior can be verified. The goal of *AbC* is to support modeling adaptive systems with the appropriate level of abstraction that permits a natural modeling and supports verification through compact models. Adaptation is guaranteed by introducing the attribute environment and its operations. However, quantitive variants of *AbC* are needed to answer questions about models' dynamics and steady-state analysis to ensure that an emergent behavior will be reached.

On the other hand, there are many calculi that aim at providing tools for specifying and reasoning about communicating systems, here we would like to touch only on those tailored for group communications while identifying the ones enjoying specific properties.

CBS [21, 24, 25] is probably the first process calculus to rely on broadcast rather than on channel-based communication. It captures the essential features of broadcast communication in a simple and natural way. Whenever a process transmits a value, all processes running in parallel and ready to input catch the broadcasted message.

The CPC calculus [11] relies on pattern-matching. Input and output prefixes are generalized to patterns whose unification enables a two-way, or symmetric, flow of information and partners are selected by matching inputs with outputs and testing for equality.

The attribute $\pi$-calculus [15, 16] aims at constraining interaction by considering values of communication attributes. A $\lambda$-function is associated to each receiving action and communication takes place only if the result of the evaluation of the function with the provided input falls within a predefined set of values.

The imperative $\pi$-calculus [14] is a recent extension of the attribute $\pi$-calculus with a global store and with imperative programs used to specify constraints.

The broadcast Quality Calculus of [29] deals with the problem of denial-of-service by means of *selective* input actions. It inspects the structure of messages by associating specific contracts to inputs, but does not provide any mean to change the input contracts during execution.

$AbC$ tries to make treasure from the experiences of the above mentioned languages and calculi in the sense that it strives for expressivity while aiming to preserve minimality and simplicity. The dynamic settings of attributes and the possibility of inspecting/modifying the environment gives $AbC$ greater flexibility and expressivity while keeping models as much natural as possible.

# 7   Concluding Remarks

We have introduced an extended and polyadic version of the $AbC$ calculus for attribute-based communication. We have investigated the expressive power of $AbC$ both in terms of its ability to model scenarios featuring collaboration, reconfiguration, and adaptation and of its ability to encode channel-based communication and other interaction paradigms. We have defined behavioral equivalences for $AbC$ and finally we have proved the correctness of the proposed encoding up to some reasonable equivalence. We demonstrated that the general concept of attribute-based communication can be exploited to provide a unifying framework to encompass different communication models. We have developed a centralized prototype implementation for $AbC$ linguistic primitives to demonstrate their simplicity and flexibility in accommodating different interaction patterns.

We plan to investigate the impact of bisimulation in terms of axioms, proof techniques, etc. for working with the calculus and to consider alternative behavioral relations like testing preorders. Further attention will be also dedicated to provide an efficient distributed implementation for $AbC$ linguistic primitives. We also plan to define a full-fledged language based on $AbC$ operators and to test its effectiveness not only as a tool for encoding calculi but also for dealing with case studies from different application domains.

## Appendix: Additional Materials

## 8   Section 3.1 Proofs

*Proof (of Lemma 1).*

The "only if" implication is straightforward. For the "if" implication, the proof is carried out by induction on the structure of $\Pi$.

– if ($\Pi = \mathtt{tt}$): according to Table6, ($\mathtt{tt} \blacktriangleright x = \mathtt{tt}$) which means that the satisfaction of $\mathtt{tt}$ does not depend on $x$ (i.e., $\Gamma \models \mathtt{tt} \blacktriangleright x$ iff $\Gamma \models \mathtt{tt}$). From Table1, We have that $\mathtt{tt}$ is satisfied by all $\Gamma$, so it is easy to that if $\Gamma \models \mathtt{tt} \blacktriangleright x$ then $\forall v. \ \Gamma[v/x] \models \mathtt{tt} \blacktriangleright x$ as required.

– if ($\Pi = \mathtt{ff}$): according to Table6, ($\mathtt{ff} \blacktriangleright x = \mathtt{ff}$) which again means that the satisfaction of $\mathtt{ff}$ does not depend on $x$. From Table1, We have that $\mathtt{ff}$ is not satisfied by any $\Gamma$, so this case holds vacuously.

– if ($\Pi = (a = m) \blacktriangleright x$): according to Table6, We have two cases:
  • if ($x = m$) then $\Pi = \mathtt{ff}$ and by induction hypotheses, the case holds vacuously.
  • if ($x \neq m$) then $\Pi = (x = m)$, according to Table1, we have that $\Gamma \models (a = m)$ iff $\Gamma(a) = m$. Since $x \neq m$, then $\Gamma(a) = m$ holds for any value of $x$ in $\Gamma$ and we have that if $\Gamma \models (a = m) \blacktriangleright x$ then $\forall v. \ \Gamma[v/x] \models (a = m) \blacktriangleright x$ as required.

– if ($\Pi = \Pi_1 \wedge \Pi_2$): according to Table6, $(\Pi_1 \wedge \Pi_2) \blacktriangleright x = (\Pi_1 \blacktriangleright x \wedge \Pi_2 \blacktriangleright x)$. From Table1, We have that $\Gamma \models (\Pi_1 \blacktriangleright x \wedge \Pi_2 \blacktriangleright x)$ iff $\Gamma \models \Pi_1 \blacktriangleright x$ and $\Gamma \models \Pi_2 \blacktriangleright x$. By induction hypotheses, We have that if ($\Gamma \models \Pi_1 \blacktriangleright x$ then $\forall v. \ \Gamma[v/x] \models \Pi_1 \blacktriangleright x$) and if ($\Gamma \models \Pi_2 \blacktriangleright x$ then $\forall v. \ \Gamma[v/x] \models \Pi_2 \blacktriangleright x$).
$\Gamma \models (\Pi_1 \blacktriangleright x \wedge \Pi_2 \blacktriangleright x)$ iff $\forall v.(\Gamma[v/x] \models \Pi_1 \blacktriangleright x \wedge \Gamma[v/x] \models \Pi_2 \blacktriangleright x)$ and now We have that if $\Gamma \models (\Pi_1 \wedge \Pi_2) \blacktriangleright x$ then $\forall v. \ \Gamma[v/x] \models (\Pi_1 \wedge \Pi_2) \blacktriangleright x$ as required.

– if ($\Pi = \Pi_1 \vee \Pi_2$): This case if analogous to the previous one.

– if ($\Pi = \neg \Pi$): According to Table6, $(\neg \Pi) \blacktriangleright x = \neg (\Pi \blacktriangleright x)$. From Table1, We have that $\Gamma \models \neg (\Pi \blacktriangleright x)$ iff not $\Gamma \models (\Pi \blacktriangleright x)$. By induction hypotheses, We have that if (not $\Gamma \models \Pi \blacktriangleright x$ then $\forall v.$ not $\Gamma[v/x] \models \Pi \blacktriangleright x$) and now We have that if $\Gamma \models \neg (\Pi) \blacktriangleright x$ then $\forall v. \ \Gamma[v/x] \models \neg (\Pi) \blacktriangleright x$ as required.

□

**Please notice that from now on and for simplifying the proofs, we use $\Pi$ to denote the semantics meaning of a predicate rather than its syntax**.

## 9   Section 5.2 Proofs

*Proof (of Lemma 2).* The proof holds by definition.                        □

The following Lemma is useful to prove that a component with a restricted name does not need any renaming when performing a $\tau$ action. We will use it in the proof of Lemma 4. .

**Lemma A1** $C[y/x] \Rightarrow C'$ *implies* $\nu x C \Rightarrow \nu y C'$

*Proof.* The proof proceeds by induction on the length of the derivation $\Rightarrow_n$

- Base Case: $n = 0$
  $C[y/x] \equiv_\alpha C'$ which implies $\nu x C \equiv_\alpha \nu y C[y/x]$ where $\equiv_\alpha$ is the structural congruence under $\alpha$-conversion.
- For all $k \leq n$:   $C[y/x] \Rightarrow_k C'$ implies $\nu x C \Rightarrow_k \nu y C'$
  if $C[y/x] \Rightarrow_{n+1} C'$, then we have that $C[y/x] \Rightarrow_n C'' \xrightarrow{\tau} C'$
  This implies that $\nu x C \Rightarrow_n \nu y C''$ and $C'' \xrightarrow{\tau} C'$ which means that $\nu y C'' \xrightarrow{\tau} \nu y C'$.
  In other words, $C'' \xrightarrow{\tau} C'$ implies $C''[y/y] \xrightarrow{\tau} C'$. Now we can apply (Res) rule. Since $y \notin fn(C'')\backslash\{y\}$ and $y \notin n(\tau)$, we have that $\nu y C'' \xrightarrow{\tau} \nu y C'$ and we have that $\nu x C \Rightarrow \nu y C'$ as required. □

*Proof (of Lemma 3).* (we only prove the weak case)
It is sufficient to prove that the relation $\mathcal{R} = \{(C_1\|C, C_2\|C)|$ for all $C$ such that $(C_1 \approx C_2)\}$ is a weak bisimulation. Depending on the last applied rule to derive the transition $C_1\|C \xrightarrow{\gamma} \hat{C}$, we have several cases.

- $C_1\|C \xrightarrow{\tau} \hat{C}$, then the last applied rule is ($\tau$-Int) or its symmetry.
  - if ($\tau$-Int) is applied then $\hat{C} = C_1'\|C$ and $C_1 \xrightarrow{\tau} C_1'$. Since $C_1 \approx C_2$ then there exists $C_2'$ such that $C_2 \Rightarrow C_2'$ and $(C_1' \approx C_2')$. By applying ($\tau$-Int) several times, we have that $C_2\|C \Rightarrow C_2'\|C$ and $(C_1'\|C, C_2'\|C) \in \mathcal{R}$
  - if the symmetry of ($\tau$-Int) is applied then $\hat{C} = C_1\|C'$ and $C \xrightarrow{\tau} C'$. So it is immediate to have that $C_2\|C \Rightarrow C_2\|C'$ and $(C_1\|C', C_2\|C') \in \mathcal{R}$
- $C_1\|C \xrightarrow{\nu\tilde{x}\overline{\Pi}\tilde{v}} \hat{C}$ with $\hat{x} \cap fn(C) = \emptyset$ and $\Pi \neq \mathsf{ff}$, then the last applied rule is (Com) or its symmetry.
  - if (Com) is applied then $\hat{C} = C_1'\|C'$, $C_1 \xrightarrow{\nu\tilde{x}\overline{\Pi}\tilde{v}} C_1'$ and $C \xrightarrow{\Pi(\tilde{v})} C'$. Since $C_1 \approx C_2$ then there exists $C_2'$ such that $C_2 \xrightarrow{\nu\tilde{x}\overline{\Pi}\tilde{v}} C_2'$ and $(C_1' \approx C_2')$. By an application of (Com) and several application of ($\tau$-Int), we have that $C_2\|C \xrightarrow{\nu\tilde{x}\overline{\Pi}\tilde{v}} C_2'\|C'$ and $(C_1'\|C', C_2'\|C') \in \mathcal{R}$
  - if the symmetry of (Com) is applied then $\hat{C} = C_1'\|C'$, $C_1 \xrightarrow{\Pi(\tilde{v})} C_1'$ and $C \xrightarrow{\nu\tilde{x}\overline{\Pi}\tilde{v}} C'$. So it is immediate to have that $C_2\|C \xrightarrow{\nu\tilde{x}\overline{\Pi}\tilde{v}} C_2'\|C'$ and $(C_1'\|C', C_2'\|C') \in \mathcal{R}$
- $C_1\|C \xrightarrow{\Pi(\tilde{v})} \hat{C}$, then the last applied rule is (Sync) and $\hat{C} = C_1'\|C'$, $C_1 \xrightarrow{\Pi(\tilde{v})} C_1'$, and $C \xrightarrow{\Pi(\tilde{v})} C'$. Since $C_1 \approx C_2$ then there exists $C_2'$ such that $C_2 \xrightarrow{\Pi(\tilde{v})} C_2'$ and $(C_1' \approx C_2')$. By an application of (Sync) and several application of ($\tau$-Int), we have that $C_2\|C \xrightarrow{\Pi(\tilde{v})} C_2'\|C'$ and $(C_1'\|C', C_2'\|C') \in \mathcal{R}$.

The strong case of bisimulation ($\sim$) follows in a similar way. □

*Proof (of Lemma 4).* (we only prove the weak case)
It is sufficient to prove that the relation $\mathcal{R} = \{(C, B)|$ $C = \nu x C_1$, $B = \nu x C_2$ with $(C_1 \approx C_2)\}$ is a weak bisimulation. We have several cases depending on the performed action in deriving the transition $C \xrightarrow{\gamma} \hat{C}$.

- if $(\gamma = \tau)$ then only rule (Res) is applied. if (Res) is applied, then $C_1[x/x] \xrightarrow{\tau} C_1'$ and $\hat{C} = \nu x C_1'$. As $(C_1 \approx C_2)$, We have that $C_2[x/x] \Rightarrow C_2'$ with $(C_1' \approx C_2')$. By Lemma A1 and several applications of (Res), we have that $B \Rightarrow \nu x C_2'$ and $(\nu x C_1', \nu x C_2') \in \mathcal{R}$.

- if $(\gamma = \nu \tilde{y} \overline{\Pi} \tilde{v})$ then either rule (Open), (Res), (Hide1) or (Hide2) is applied.
    - if (Open) is applied, then $x \in (\tilde{v} \cup \tilde{y}) \backslash n(\Pi)$ and $C_1[z/x] \xrightarrow{\overline{\Pi} \tilde{v}} C_1'$ with $\hat{C} = C_1'$. As $(C_1 \approx C_2)$, We have that $C_2[z/x] \xrightarrow{\overline{\Pi} \tilde{v}} C_2'$ with $(C_1' \approx C_2')$. By Lemma A1, an application of (Open), and several applications of (Res), we have that $B \xrightarrow{\nu \tilde{y} \overline{\Pi} \tilde{v}} C_2'$ and $(C_1', C_2') \in \mathcal{R}$.
    - if (Res) is applied, then $C_1[z/x] \xrightarrow{\nu \tilde{y} \overline{\Pi} \tilde{v}} C_1'$ and $\hat{C} = \nu z C_1'$. As $(C_1 \approx C_2)$, We have that $C_2[z/x] \xrightarrow{\nu \tilde{y} \overline{\Pi} \tilde{v}} C_2'$ with $(C_1' \approx C_2')$. By Lemma A1 and several applications of (Res), we have that $B \xrightarrow{\nu \tilde{y} \overline{\Pi} \tilde{v}} \nu z C_2'$ and $(\nu z C_1', \nu z C_2') \in \mathcal{R}$
    - if (Hide1) is applied, then $C_1 \xrightarrow{\nu \tilde{y} \overline{\Pi} \tilde{v}} C_1'$ and $\hat{C} = \nu x \nu \tilde{y} C_1'$. As $(C_1 \approx C_2)$, We have that $C_2 \xrightarrow{\nu \tilde{y} \overline{\Pi} \tilde{v}} C_2'$ with $(C_1' \approx C_2')$. By Lemma A1, an application of (EHide1), and several applications of (Res), we have that $B \xrightarrow{\nu \tilde{y} \overline{\overline{\Pi}} \tilde{v}} \nu x \nu \tilde{y} C_2'$ and $(\nu x \nu \tilde{y} C_1', \nu x \nu \tilde{y} C_2') \in \mathcal{R}$
    - if (Hide2) is applied, then $C_1 \xrightarrow{\nu \tilde{y} \overline{\Pi} \tilde{v}} C_1'$ and $\hat{C} = \nu x C_1'$. As $(C_1 \approx C_2)$, We have that $C_2 \xrightarrow{\nu \tilde{y} \overline{\Pi} \tilde{v}} C_2'$ with $(C_1' \approx C_2')$. By Lemma A1, an application of (EHide2), and several applications of (Res), we have that $B \xrightarrow{\nu \tilde{y} \overline{\Pi} \blacktriangleright \overline{x} \tilde{v}} \nu x C_2'$ and $(\nu x C_1', \nu x C_2') \in \mathcal{R}$
- if $(\gamma = \Pi(\tilde{v}))$ then $x \notin n(\gamma)$ and only rule (Res) is applied. So we have that $C_1[y/x] \xrightarrow{\Pi(\tilde{v})} C_1'$ and $\hat{C} = \nu y C_1'$. As $(C_1 \approx C_2)$, We have that $C_2[y/x] \xrightarrow{\Pi(\tilde{v})} C_2'$ with $(C_1' \approx C_2')$. By Lemma A1 and several applications of (Res), we have that $B \xrightarrow{\Pi(\tilde{v})} \nu y C_2'$ and $(\nu y C_1', \nu y C_2') \in \mathcal{R}$

The strong case of bisimulation ($\sim$) follows in a similar way. □

*Proof (of Lemma 5).* (we only prove the weak case)
It is sufficient to prove that the relation $\mathcal{R} = \{(C, B) | \ C = !C_1, \ B = !C_2$ with $(C_1 \approx C_2)\}$ is a weak bisimulation. The proof follows easily by applying rule (Rep). if $C \xrightarrow{\gamma} \hat{C}$, so we have that $C_1 \xrightarrow{\gamma} C_1'$ and $\hat{C} = C_1' \| !C_1$. As $(C_1 \approx C_2)$, , then there exists $C_2'$ such that $C_2 \xrightarrow{\gamma} C_2'$ with $(C_1' \approx C_2')$. By an application of rule (Rep) and several applications of rule (Comp), we have that $B \xrightarrow{\gamma} C_2' \| !C_2$ and $(C_1' \| !C_1, \ C_2' \| !C_2) \in \mathcal{R}$ as required.

The strong case of bisimulation ($\sim$) follows in a similar way. □

*Proof (of Theorem 1).* (we only prove the weak case)
It is sufficient to prove that bisimilarity is barb-preserving, reduction-closed, and context-closed.

- (Barb-preservation): By the definition of the barb $C_1{\downarrow}_\Pi$ if $C_1 \xrightarrow{\nu\tilde{x}\overline{\Pi}\tilde{v}}$ for an output label $\nu\tilde{x}\overline{\Pi}\tilde{v}$ with $\Pi \neq \mathsf{ff}$. As $(C_1 \approx C_2)$, We have that also $C_2 \xLongrightarrow{\nu\tilde{x}\overline{\Pi}\tilde{v}}$ and $C_2 {\Downarrow}_\Pi$.

- (Reduction-closure): $C_1 \rightarrowtail C_1'$ means that either $C_1 \xrightarrow{\tau} C_1'$ or $C_1 \xrightarrow{\nu\tilde{x}\overline{\Pi}\tilde{v}} C_1'$. As $(C_1 \approx C_2)$, then there exists $C_2'$ such that either $C_2 \Rightarrow C_2'$ or $C_2 \xLongrightarrow{\nu\tilde{x}\overline{\Pi}\tilde{v}} C_2'$ with $(C_1' \approx C_2')$. So $C_2 \rightarrowtail^* C_2'$.

- (Context-closure): Let $(C_1 \approx C_2)$ and let $\mathcal{C}[\bullet]$ be an arbitrary AbC-context. By induction on the structure of $\mathcal{C}[\bullet]$ and using Lemma 3, Lemma 4, and Lemma 5, We have that $\mathcal{C}[c_1] \approx \mathcal{C}[c_2]$.

In conclusion, We have that $(C_1 \cong C_2)$ as required. $\qquad\square$

*Proof (of Lemma 6).* (we only prove the weak case)

It is sufficient to prove that the relation $\mathcal{R} = \{(C_1, C_2) \mid C_1 \cong C_2\}$ is a weak bisimulation.

1. Suppose that $C_1 \xrightarrow{\nu\tilde{x}\overline{\Pi}\tilde{v}} C_1'$ for any $\Pi$ and a sequence of values $\tilde{v}$ where $\Pi \neq \mathsf{ff}$. We build up a context to mimic the effect of this transition. Our context has the following form:

$$\mathcal{C}[\bullet] \triangleq [\bullet] \parallel \prod_{i \in I} \Gamma_i : \Pi_i(\tilde{x}_i).\langle \tilde{x}_i = \tilde{v}\rangle(\tilde{x}_i,\ a)@(in = a)$$
$$\parallel \prod_{j \in J} \Gamma_j : (y = a)(\tilde{x}_j,\ y).(\tilde{x}_j,\ b)@(out = b)$$

where $|\tilde{x}_i| = |\tilde{x}_j|$, $I \cap J = \emptyset$ and $\Gamma_j \models (in = a)$, and the names $a$ and $b$ are fresh. $\Pi_i$ is an arbitrary predicate. We use the notation $\langle \tilde{x}_i = \tilde{v}\rangle$ to denote $\langle (x_{i,1} = v_1) \wedge (x_{i,2} = v_2) \wedge \cdots \wedge (x_{i,n} = v_n)\rangle$ where $n = |\tilde{x}_i|$ and $\prod_{i \in I} \Gamma_i : P_i$ to denote the parallel composition of all components $\Gamma_i : P_i$, for $i \in I$. Now assume that $(\Gamma_i \models \Pi)$ and $\Pi_i$ is satisfied given the sequence of values $\tilde{v}$. Intuitively, the existence of a barb on $(in = a)$ indicates that the action has not yet happened, whereas the presence of a barb on $(out = b)$ together with the absence of the barb on $(in = a)$ ensures that the action happened.

As $\cong$ is context-closed, $C_1 \cong C_2$ implies $\mathcal{C}[C_1] \cong \mathcal{C}[C_2]$. Since $C_1 \xrightarrow{\nu\tilde{x}\overline{\Pi}\tilde{v}} C_1'$, it follows by Lemma 2 that:

$$\mathcal{C}[C_1] \Rightarrow C_1' \parallel \prod_{i \in I} \Gamma_i : 0 \parallel \prod_{j \in J} \Gamma_j : (\tilde{v},\ b)@(out = b) = \hat{C}_1$$

with $\hat{C}_1 \not\Downarrow_{(in=a)}$ and $\hat{C}_1 \Downarrow_{(out=b)}$.

The reduction sequence above must be matched by a corresponding reduction sequence $\mathcal{C}[C_2] \Rightarrow \hat{C}_2 \cong \hat{C}_1$ with $\hat{C}_2 \not\Downarrow_{(in=a)}$ and $\hat{C}_2 \Downarrow_{(out=b)}$. The conditions on the barbs allow us to get the structure of the above reduction sequence as follows:

$$\mathcal{C}[C_2] \Rightarrow C_2' \parallel \prod_{i \in I} \Gamma_i : 0 \parallel \prod_{j \in J} \Gamma_j : (\tilde{v},\ b)@(out = b) \cong \hat{C}_1$$

This implies that $C_2 \xrightarrow{\nu \tilde{x} \overline{\Pi} \tilde{v}} C_2'$. Reduction barbed congruence is preserved by name restriction, so we have that $\nu a \nu b \hat{C}_1 \cong \nu a \nu b \hat{C}_2$ and $C_1' \cong C_2'$ as required.

2. Suppose that $C_1 \xrightarrow{\Pi(\tilde{v})} C_1'$ for any $\Pi$ and a sequence of values $\tilde{v}$. Assume $C_1 \equiv \Gamma : P_1$, we build up the following context to mimic the effect of this transition.

$$\mathcal{C}[\bullet] \triangleq [\bullet] \parallel \Gamma' : (\tilde{v})@(in = a).(\tilde{v})@(out = b)$$

where $\Gamma \models \Pi$ and $\Pi = (in = a)$, and the names $a$ and $b$ are fresh. As $\cong$ is context-closed, $C_1 \cong C_2$ implies $\mathcal{C}[C_1] \cong \mathcal{C}[C_2]$. Since $C_1 \xrightarrow{\Pi(\tilde{v})} C_1'$, it follows by Lemma 2 that:

$$\mathcal{C}[C_1] \Rightarrow C_1' \parallel (\tilde{v})@(out = b) = \hat{C}_1$$

with $\hat{C}_1 \not\Downarrow_{(in=a)}$ and $\hat{C}_1 \Downarrow_{(out=b)}$.

The reduction sequence above must be matched by a corresponding reduction sequence $\mathcal{C}[C_2] \Rightarrow \hat{C}_2 \cong \hat{C}_1$ with $\hat{C}_2 \not\Downarrow_{(in=a)}$ and $\hat{C}_2 \Downarrow_{(out=b)}$ as follows:

$$\mathcal{C}[C_2] \Rightarrow C_2' \parallel (\tilde{v})@(out = b) \cong \hat{C}_1$$

This implies that $C_2 \xRightarrow{\Pi(\tilde{v})} C_2'$. Reduction barbed congruence is preserved by name restriction, so we have that $\nu a \nu b \hat{C}_1 \cong \nu a \nu b \hat{C}_2$ and $C_1' \cong C_2'$ as required.

3. Suppose that $C_1 \xrightarrow{\tau} C_1'$. This case is straightforward.    □

### 9.1   Section 5.3 Proofs

*Proof (of Lemma 7).* The proof proceeds by induction on the shortest transition of $\rightarrow_{b\pi}$. We have several cases depending on the structure of the term $P$.

– if $P \triangleq \mathtt{nil}$: This case is immediate $(\!|\mathtt{nil}|\!)_c \triangleq \emptyset : 0$

– if $P \triangleq \tau.G$: We have that $\tau.G \xrightarrow{\tau} G$ and it is translated to $(\!|\tau.G|\!)_c \triangleq \emptyset : ()@\mathsf{ff}.(\!|G|\!)_p$. We can only apply rule (Comp) to mimic this transition.

$$\emptyset : ()@\mathsf{ff}.(\!|G|\!)_p \xrightarrow{\overline{\mathsf{ff}}()} \emptyset : (\!|G|\!)_p$$

$$\emptyset : ()@\mathsf{ff}.(\!|G|\!)_p \xrightarrow{\overline{\mathsf{ff}}()} \emptyset : (\!|G|\!)_p$$

Now it is not hard to see that $(\!|\ G\ |\!)_c \simeq \emptyset : (\!|G|\!)_p$. They are even structural congruent. Notice that sending on a false predicate is not observable (i.e., a silent move).

– if $P \triangleq a(\tilde{x}).G$: We have that $a(\tilde{x}).G \xrightarrow{a(\tilde{z})} G[\tilde{z}/\tilde{x}]$ and it is translated to $(\!|a(\tilde{x}).Q|\!)_c \triangleq \emptyset : \Pi(y,\tilde{x}).(\!|G|\!)_p$ where $\Pi = (y = a)$. We can only apply rule (Comp) to mimic this transition.

$$\dfrac{\emptyset : \Pi(y,\tilde{x}).(\!|G|\!)_p \xrightarrow{(a=a)(a,\ \tilde{z})} \emptyset : (\!|G|\!)_p[a/y,\ \tilde{z}/\tilde{x}]}{\emptyset : \Pi(y,\tilde{x}).(\!|G|\!)_p \xrightarrow{(a=a)(a,\ \tilde{z})} \emptyset : (\!|G|\!)_p[a/y,\ \tilde{z}/\tilde{x}]}$$

It is not hard to see that: $(\!|G[\tilde{z}/\tilde{x}]|\!)_c \simeq \emptyset : (\!|G|\!)_p[a/y,\ \tilde{z}/\tilde{x}] \simeq \emptyset : (\!|G|\!)_p[\tilde{z}/\tilde{x}]$ since $y \notin n((\!|G|\!)_p)$.

– if $P \triangleq \bar{a}\tilde{x}.G$: The proof is similar to the previous case but by applying this output transition instead.

– The fail rules for **nil**, $\tau$, input and output are proved in a similar way but with applying (C-Fail) instead.

– if $P \triangleq \nu x Q$: We have that either $\nu x Q \xrightarrow{\gamma} \nu x Q'$ , $\nu x Q \xrightarrow{\tau} \nu x \nu \tilde{y} Q'$ or $\nu x Q \xrightarrow{\nu x \nu \tilde{y} \bar{a} \tilde{z}} Q'$ and it is translated to $(\!|\nu x Q|\!)_c \triangleq \nu x \emptyset : (\!|Q|\!)_p$. We prove each case independently.

  • Case $\nu x Q \xrightarrow{\gamma} \nu x Q'$ : By applying induction hypotheses on the premise $Q \xrightarrow{\gamma} Q'$, we have that $(\!|Q|\!)_c \twoheadrightarrow^* \simeq (\!|Q'|\!)_c$. We can only use rule (Res) to mimic transition depending on the performed action.

$$\dfrac{\emptyset : (\!|Q|\!)_p[y/x] \xrightarrow{\gamma} \emptyset : (\!|Q'|\!)_p[y/x]}{\nu x \emptyset : (\!|Q|\!)_p \xrightarrow{\gamma} \nu y \emptyset : (\!|Q'|\!)_p[y/x]}$$

  And we have that $(\!|\nu x Q'|\!)_c \simeq \nu y \emptyset : (\!|Q'|\!)_p[y/x]$ as required.

  • Case $\nu a Q \xrightarrow{\tau} \nu a \nu \tilde{y} Q'$ : By applying induction hypotheses on the premise $Q \xrightarrow{\nu \tilde{y} \bar{a} \tilde{z}} Q'$, we have that $(\!|Q|\!)_c \twoheadrightarrow^* \simeq (\!|Q'|\!)_c$. We can only use (Hide1) to mimic this transition.

$$\dfrac{\emptyset : (\!|Q|\!)_p \xrightarrow{\nu \tilde{y} a = \bar{a}(a,\ \tilde{z})} \emptyset : (\!|Q'|\!)_p}{\nu a \emptyset : (\!|Q|\!)_p \xrightarrow{\nu \tilde{y} \overline{\overline{a}}(a,\ \tilde{z})} \nu a \nu \tilde{y} \emptyset : (\!|Q'|\!)_p}$$

  We have that $(\!|\nu a \nu \tilde{y} Q'|\!)_c \simeq \nu x \nu \tilde{y} \emptyset : (\!|Q'|\!)_p$ as required.

  • Case $\nu x Q \xrightarrow{\nu x \nu \tilde{y} \bar{a} \tilde{z}} Q'$: follows in a similar way using rule (Open) .
  • Case $\nu x Q \xrightarrow{\alpha;}$: is similar to the case with (Res) rule.

– if $P \triangleq ((rec\ A\langle \tilde{x} \rangle).P)\langle \tilde{y} \rangle)$: This case is trivial.

– if $P \triangleq G_1 + G_2$: We have that either $G_1 + G_2 \xrightarrow{\alpha} G_1'$ or $G_1 + G_2 \xrightarrow{\alpha} G_2'$. We only consider the first case with $G_1 \xrightarrow{\alpha} G_1'$ and the other case follows in a similar way. This process is translated to $(\!|G_1 + G_2|\!)_c \triangleq \emptyset : (\!|G_1|\!)_p + (\!|G_2|\!)_p$. By applying induction hypotheses on the premise $G_1 \xrightarrow{\alpha} G_1'$, we have that $(\!|\ G_1\ |\!)_c \twoheadrightarrow^* \simeq (\!|\ G_1'\ |\!)_c$. We can apply either rule (Comp) or rule (C-Fail) (i.e., when discarding) to mimic this transition depending on the performed action.

We consider the case of (Comp) only and the other case follows in a similar way.

$$\frac{\emptyset : (\!|G_1|\!)_p \overset{\lambda}{\longmapsto} \emptyset : (\!|G_1'|\!)_p}{\dfrac{\emptyset : (\!|G_1|\!)_p + (\!|G_2|\!)_p \overset{\lambda}{\longmapsto} \emptyset : (\!|G_1'|\!)_p}{\emptyset : (\!|G_1|\!)_p + (\!|G_2|\!)_p \overset{\gamma}{\longrightarrow} \emptyset : (\!|G_1'|\!)_p}}$$

Again $(\!|G_1'|\!)_c \simeq \emptyset : (\!|G_1'|\!)_p$

- if $P \triangleq P_1 \| P_2$: This process is translated to $(\!| P_1 \| P_2 |\!)_c \triangleq \emptyset : (\!| P_1 |\!)_p \| \emptyset : (\!| P_2 |\!)_p$. We have four cases depending on the performed action in deriving the transition $P_1 \| P_2 \overset{\alpha}{\longrightarrow} \tilde{P}$.

  • $P_1 \| P_2 \xrightarrow{\nu\tilde{y}\tilde{a}\tilde{x}} P_1' \| P_2'$: We have two cases, either $P_1 \xrightarrow{\nu\tilde{y}\tilde{a}\tilde{x}} P_1'$ and $P_2 \xrightarrow{a(\tilde{x})} P_2'$ or $P_2 \xrightarrow{\nu\tilde{y}\tilde{a}\tilde{x}} P_2'$ and $P_1 \xrightarrow{a(\tilde{x})} P_1'$. We only consider the first case and the other case follows in the same way. By applying induction hypotheses on the premises $P_1 \xrightarrow{\nu\tilde{y}\tilde{a}\tilde{x}} P_1'$ and $P_2 \xrightarrow{a(\tilde{x})} P_2'$, we have that $(\!|P_1|\!)_c \twoheadrightarrow^* \simeq (\!|P_1'|\!)_c$ and $(\!|P_2|\!)_c \twoheadrightarrow^* \simeq (\!|P_2'|\!)_c$. We only can apply (Com).

$$\frac{\emptyset : (\!|P_1|\!)_p \xrightarrow{\nu\tilde{y}\overline{(a=a)}(a,\tilde{x})} \emptyset : (\!|P_1'|\!)_p \qquad \emptyset : (\!|P_2|\!)_p \xrightarrow{(a=a)(a,\tilde{x})} \emptyset : (\!|P_2'|\!)_p}{\emptyset : (\!|P_1|\!)_p \parallel \emptyset : (\!|P_2|\!)_p \xrightarrow{\nu\tilde{y}\overline{(a=a)}(a,\tilde{x})} \emptyset : (\!|P_1'|\!)_p \parallel \emptyset : (\!|P_2'|\!)_p}$$

Again we have that: $(\!|P_1' \| P_2'|\!)_c \simeq \emptyset : (\!|P_1'|\!)_p \| \emptyset : (\!|P_2'|\!)_p$. Notice that the $b\pi$ term and its encoding have the same observable behavior i.e., $P_1 \| P_2 \downarrow_a$ and $(\!|P_1 \| P_2|\!)_c \downarrow_{(a=a)}$.

  • $P_1 \| P_2 \xrightarrow{a(\tilde{x})} P_1' \| P_2'$: By applying induction hypotheses on the premises $P_1 \xrightarrow{a(\tilde{x})} P_1'$ and $P_2 \xrightarrow{a(\tilde{x})} P_2'$, we have that $(\!|P_1|\!)_c \twoheadrightarrow^* \simeq (\!|P_1'|\!)_c$ and $(\!|P_2|\!)_c \twoheadrightarrow^* \simeq (\!|P_2'|\!)_c$. We only can apply (Sync) to mimic this transition.

$$\frac{\emptyset : (\!|P_1|\!)_p \xrightarrow{(a=a)(a,\tilde{x})} \emptyset : (\!|P_1'|\!)_p \qquad \emptyset : (\!|P_2|\!)_p \xrightarrow{(a=a)(a,\tilde{x})} \emptyset : (\!|P_2'|\!)_p}{\emptyset : (\!|P_1|\!)_p \parallel \emptyset : (\!|P_2|\!)_p \xrightarrow{(a=a)(a,\tilde{x})} \emptyset : (\!|P_1'|\!)_p \parallel \emptyset : (\!|P_2'|\!)_p}$$

Again we have that: $(\!|P_1' \| P_2'|\!)_c \simeq \emptyset : (\!|P_1'|\!)_p \| \emptyset : (\!|P_2'|\!)_p$.

  • $P_1 \| P_2 \overset{\alpha}{\longrightarrow} P_1' \| P_2$ if $P_1 \overset{\alpha}{\longrightarrow} P_1'$ and $P_2 \xrightarrow{sub(\alpha):}$ or $P_1 \| P_2 \overset{\alpha}{\longrightarrow} P_1 \| P_2'$ if $P_2 \overset{\alpha}{\longrightarrow} P_2'$ and $P_1 \xrightarrow{sub(\alpha):}$. we consider only the first case and by applying induction hypotheses on the premises $P_1 \overset{\alpha}{\longrightarrow} P_1'$ and $P_2 \xrightarrow{sub(\alpha):}$, we have that $(\!|P_1|\!)_c \twoheadrightarrow^* \simeq (\!|P_1'|\!)_c$ and $(\!|P_2|\!)_c \twoheadrightarrow^* \simeq (\!|P_2|\!)_c$. We have many cases depending on the performed action:

    * if $\alpha = \tau$ then $P_1 \| P_2 \overset{\tau}{\longrightarrow} P_1' \| P_2$ with $P_1 \overset{\tau}{\longrightarrow} P_1'$ and $P_2 \xrightarrow{sub(\tau):}$. We can apply ($\tau$-Int) to mimic this transition.

$$\frac{\emptyset : (\!|P_1|\!)_p \xrightarrow{\nu\tilde{y}\overline{\Pi}\tilde{x}} \emptyset : (\!|P_1'|\!)_p \qquad \Pi \simeq \mathsf{ff}}{\emptyset : (\!|P_1|\!)_p \| \emptyset : (\!|P_2|\!)_p \overset{\tau}{\longrightarrow} \emptyset : (\!|P_1'|\!)_p \| \emptyset : (\!|P_2|\!)_p}$$

and again we have that: $(\!|P_1'\|P_2|\!)_c \simeq \emptyset : (\!|P_1'|\!)_p \| \emptyset : (\!|P_2|\!)_p$.

∗ if $\alpha = a(\tilde{x})$: then $P_1\|P_2 \xrightarrow{a(\tilde{x})} P_1'\|P_2$ with $P_1 \xrightarrow{a(\tilde{x})} P_1'$ and $P_2 \xrightarrow{a:}$ . We can apply (Sync) to mimic this transition.

$$\frac{\emptyset : (\!|P_1|\!)_p \xrightarrow{(a=a)(a,\tilde{x})} \emptyset : (\!|P_1'|\!)_p \quad \dfrac{\emptyset : (\!|P_2|\!)_p \xrightarrow{\widetilde{(a=a)(a,\tilde{x})}} \emptyset : (\!|P_2|\!)_p}{\emptyset : (\!|P_2|\!)_p \xrightarrow{(a=a)(a,\tilde{x})} \emptyset : (\!|P_2|\!)_p} \text{ C-Fail}}{\emptyset : (\!|P_1|\!)_p \| \emptyset : (\!|P_2|\!)_p \xrightarrow{(a=a)(a,\tilde{x})} \emptyset : (\!|P_1'|\!)_p \| \emptyset : (\!|P_2|\!)_p}$$

Again we have that: $(\!|P_1'\|P_2|\!)_c \simeq \emptyset : (\!|P_1'|\!)_p \| \emptyset : (\!|P_2|\!)_p$.

∗ if $\alpha = \nu\tilde{y}\bar{a}\tilde{x}$ then $P_1\|P_2 \xrightarrow{\nu\tilde{y}\bar{a}\tilde{x}} P_1'\|P_2$ with $P_1 \xrightarrow{\nu\tilde{y}\bar{a}\tilde{x}} P_1'$ and $P_2 \xrightarrow{a:}$. We can apply (Comp).

$$\frac{\emptyset : (\!|P_1|\!)_p \xrightarrow{\nu\tilde{y}\overline{(a=a)}(a,\tilde{x})} \emptyset : (\!|P_1'|\!)_p \quad \dfrac{\emptyset : (\!|P_2|\!)_p \xrightarrow{\widetilde{(a=a)(a,\tilde{x})}} \emptyset : (\!|P_2|\!)_p}{\emptyset : (\!|P_2|\!)_p \xrightarrow{(a=a)(a,\tilde{x})} \emptyset : (\!|P_2|\!)_p} \text{ C-Fail}}{\emptyset : (\!|P_1|\!)_p \| \emptyset : (\!|P_2|\!)_p \xrightarrow{\nu\tilde{y}\overline{(a=a)}(a,\tilde{x})} \emptyset : (\!|P_1'|\!)_p \| \emptyset : (\!|P_2|\!)_p}$$

Again we have that: $(\!|P_1'\|P_2|\!)_c \simeq \emptyset : (\!|P_1'|\!)_p \| \emptyset : (\!|P_2|\!)_p$. Notice that the $b\pi$ term and its encoding have the same observable behavior i.e., $P_1\|P_2 \downarrow_a$ and $(\!|P_1\|P_2|\!) \downarrow_{(a=a)}$.

□

*Proof (of Lemma 8).* The proof holds immediately due to the fact that every encoded $b\pi$-term (i.e., $(\!| P |\!)_\emptyset$) has exactly one possible transition which matches the original $b\pi$-term (i.e., $P$).